\documentclass[usegraphicx,useAMS,usenatbib]{mn2e}
\usepackage{times}
\usepackage{amssymb}
\usepackage{amsmath}
\usepackage{graphicx}
\usepackage{euscript}
\usepackage{rotating}
\usepackage{epsfig}
\usepackage{mathptmx}
\usepackage{amsbsy}

\def\arc{{\rm\thinspace arcsec}}
\def\cm{{\rm\thinspace cm}}

\def\erg{{\rm\thinspace erg}}

\def\ha{H$\alpha$}

\def\km{{\rm\thinspace km}}
\def\kpc{{\rm\thinspace kpc}}

\def\Msun{\hbox{$\rm\thinspace M_{\odot}$}}
\def\Zsun{\hbox{$\rm\thinspace Z_{\odot}$}}

\def\s{{\rm\thinspace s}}
\def\yr{{\rm\thinspace yr}}
\def\Hz{{\rm\thinspace Hz}}

\def\pHz{\hbox{$\Hz^{-1}\,$}}

\def\ergpspcmsqphz{\hbox{$\erg\cm^{-2}\s^{-1}\pHz\,$}}

\def\ergpsphz{\hbox{$\erg\s^{-1}\pHz\,$}}

\def\ergpcmsqpspsqarcsec{\hbox{$\erg\cm^{-2}\s^{-1}\arc^{-2}\,$}}

\def\ergps{\hbox{$\erg\s^{-1}\,$}}

\def\kmps{\hbox{$\km\s^{-1}\,$}}

\def\Msunpyr{\hbox{$\Msun\yr^{-1}\,$}}

\def\pHz{\hbox{$\Hz^{-1}\,$}}

\def\Msunpyrpsqkpc{\hbox{$\Msunpyr\kpc^{-2}$}}
\def\gband{$g_{475}$}
\def\iband{$I_{814}$}
\def\lya{Ly$\alpha$}

\DeclareMathAlphabet{\vib}{OML}{cmm}{m}{it}

\begin{document}
\include{defn} \title{Diffuse UV light associated with the Spiderweb Galaxy: evidence for in-situ star formation outside galaxies\\}
\author[N.~A.~Hatch, R.~A.~Overzier, H.~J.~A.~R\"ottgering, J.~D.~Kurk, and G.~K.~Miley]
{N. A. Hatch$^{1}$\thanks{E-mail:hatch@strw.leidenuniv.nl}, R.~A.~Overzier$^{2}$, H.~J.~A.~R\"ottgering$^{1}$, J.~D.~Kurk$^{3}$, and G.~K.~Miley$^{1}$ \\$^{1}$Leiden Observatory, University of Leiden, P.B. 9513, Leiden 2300 RA, The Netherlands\\$^{2}$The Johns Hopkins University, 3400 N. Charles Street, Baltimore, MD 21218, USA\\$^{3}$Max-Planck-Institute f\"ur Astronomie, K\"onigstuhl 17, 69117 Heidelberg, Germany
}\maketitle
\begin{abstract} We present detailed images of diffuse UV intergalactic light (IGL), situated in a 60\,kpc halo that surrounds the radio galaxy MRC\,1138-262 at z=2. We discuss the nature of the IGL and rule out faint cluster galaxies, nebular continuum emission, synchrotron, inverse Compton, synchrotron self-Compton emission and scattering of galactic stellar light as possible sources of the IGL. Dust scattered quasar light is an unlikely possibility that cannot be ruled out entirely. We conclude that the source of the IGL is most likely to be a young stellar population distributed in a halo encompassing the radio and satellite galaxies, undergoing  star formation at a rate greater than 57$\pm8$\Msunpyr. Within 70\,kpc of the radio core, approximately 44\% of the star formation that is traced by UV light occurs in this diffuse mode. The average  UV colour of the IGL is bluer than the average galaxy colour, and there is a trend for the IGL to become bluer with increasing radius from the radio galaxy. Both the galaxies and the IGL show a UV colour--surface brightness relation which can be obtained by variations in either stellar population age or extinction. These observations show a different, but potentially important mode of star formation, that is diffuse in nature. Star formation, as traced by UV light,  occurs in two modes in the high redshift universe: one in the usual Lyman break galaxy clump-like mode on kpc scales, and the other in a diffuse mode over a large region surrounding massive growing galaxies.  Such a mode of star formation can easily be missed by high angular resolution observations that are well suited for detecting high surface brightness compact galaxies. Extrapolating from these results, it is possible that a significant amount of star formation occurs in large extended regions within the halos of the most massive galaxies forming at high redshift.
\end{abstract}
\section{Introduction}
Distant powerful radio galaxies are important probes of the formation and evolution of massive galaxies, because they are among the most luminous and massive galaxies known in the early Universe, and are likely progenitors of dominant cluster galaxies \cite[e.g][]{Pentericci1997,Villar-Martin2006,Seymour2007}. They are generally embedded in giant (cD-sized) ionized gas halos \cite[e.g][]{Reuland2003} and surrounded by galaxy overdensities \citep{Pentericci2000,Venemans2007}. With radio lifetimes (few $\times10^7$yr) being much smaller than cosmological timescales, the statistics are consistent with every dominant cluster galaxy having gone through a luminous radio phase during its evolution.

The deepest image of a distant radio galaxy with the {\it Hubble Space Telescope (HST)} is a recent {\it Advanced Camera for Surveys (ACS)} observation of MRC\,1138-262 at z$ = 2.156$ \citep{Miley2006}. This system has been dubbed the Spiderweb galaxy as its complex morphology resembles a spider's web.  The {\it ACS} image shown in Fig.~1 shows tens of UV bright galaxies surrounding a central galaxy (marked with a white cross). As the radio core is cospatial with the central galaxy we refer to this central galaxy as the radio galaxy, and all the surrounding galaxies are referred to as the satellite galaxies. All of these galaxies may merge together to form a single massive galaxy at z=0, therefore the whole complex system is referred to as the Spiderweb system This system is one of the most intensively studied distant radio galaxies \citep{Pentericci1997,Pentericci1998,Pentericci2000} and exhibits the properties expected for a progenitor of a dominant cluster galaxy. The infrared luminosity provides an upper limit to the stellar mass of 10$^{12}$\Msun\ \citep{Seymour2007}, so it is one of the most massive galaxies known at z$>$2. The host galaxy is surrounded by a giant Ly$\alpha$ halo and the high rotation measure of the radio source means that it is embedded in a dense medium with an ordered magnetic field \citep{Pentericci1997}. The radio galaxy is associated with a $>$3\,Mpc-sized structure of galaxies with an estimated mass $>4\times10^{14}$\Msun\ \citep{Venemans2007}, indicating that it is the predecessor of a local rich cluster. Mechanical feedback from the active galactic nucleus (AGN) is sufficient to expel significant fractions of the interstellar medium of the massive gas-rich galaxy  \citep{Nesvadba2006}. Therefore the AGN may quench the star formation and allow the radio galaxy to evolve onto the red sequence as suggested by popular models of massive galaxy formation \citep{Croton2006}. The study of the formation of the most massive galaxies through these and other mechanisms has so far mainly been limited to models and simulations \cite[e.g][]{Dubinski1998,Gao2004,DeLucia2007}. Our aim is to use these observations of a high-redshift radio galaxy to study brightest cluster galaxy formation.

In this work we present detailed images of widespread UV intergalactic light (IGL) situated in a 60\,kpc halo that lies between the radio galaxy and the UV bright satellite galaxies. We examine the possible origins of this light and present in-situ star formation as the most probable. In section \ref{method} we describe the observations and data reduction, in section \ref{results} we describe the distribution and colour of the IGL, and discuss possible origins in section \ref{nature}. Section \ref{discussion} discusses the implications of widespread star formation in a halo around a massive forming brightest cluster galaxy. Throughout this work we use $H_0=71$, $\Omega_M=0.27$, and $\Omega_\Lambda=0.73$ \citep{Spergel2003}. All magnitudes are AB magnitudes. At a redshift of 2.156, the linear scale is 8.4\,kpc/".
\section{Method}
\label{method}
\subsection{Observations}
\begin{figure*}
\centering
\includegraphics[width=0.9\columnwidth]{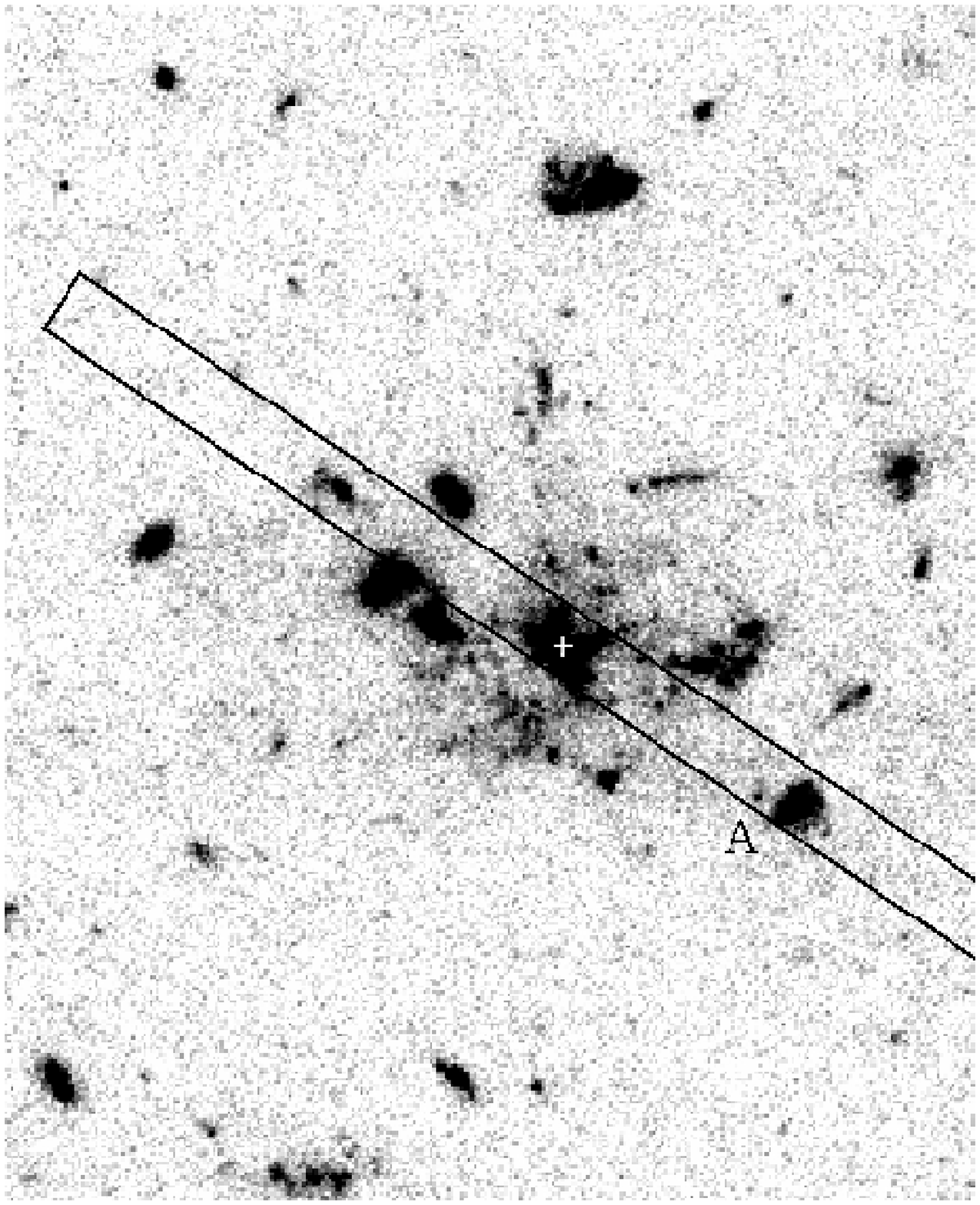}
\includegraphics[width=0.9\columnwidth]{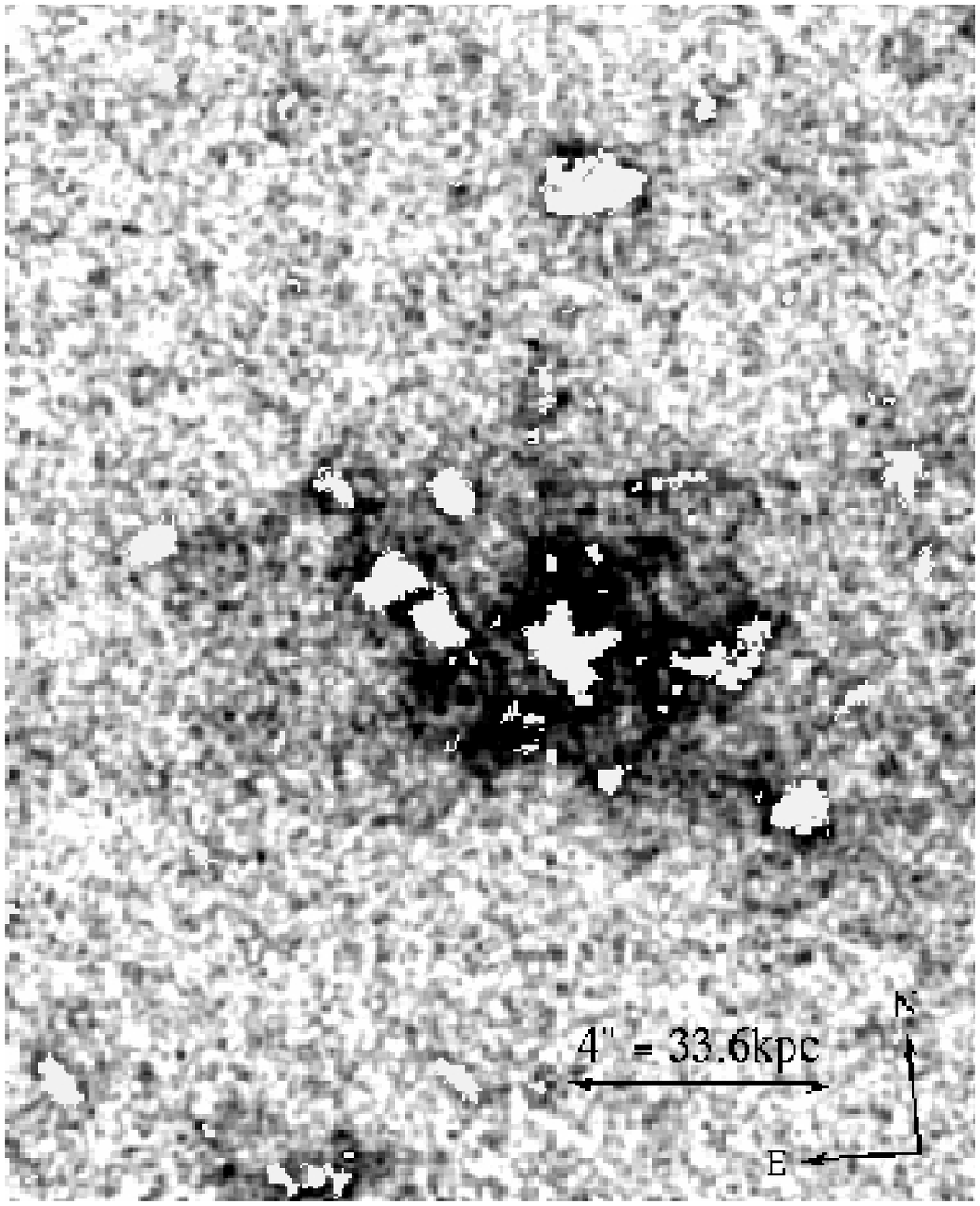}
\caption{{\bf Left}:  Combined \gband$+$\iband\ image of the central 15"$\times$18.5" region  of the Spiderweb complex. The radio galaxy, MRC\,1138-262, is marked with a white cross, and is surrounded by numerous satellite galaxies. The black box indicates the position of the MOS slit from which the spectra in Figure \ref{spectra} are extracted. The label A indicates the galaxy A whose spectrum is shown in Figure \ref{spectra}. {\bf Right}: \gband\ image of MRC\,1138-262  and surrounding 15"$\times$18.5" region, smoothed with a gaussian of $\sigma=0.05$\,arcsec. All regions identified as \lq galaxies' by {\sc Sextractor} have been removed and are coloured white. The intergalactic light is clearly visible between the radio and satellite galaxies.\label{image}}
\end{figure*}

The imaging data used in this work was obtained with the {\it ACS} on the {\it HST}. The total exposure time in the F475W ($g_{475}$) filter was 9 orbits (20670\,s) and 10 orbits (23004\,s) in the F814W ($I_{814}$) filter.  The data were reduced using the ACS pipeline science investigation software ({\it Apsis}; \citealt{Blakeslee2003}). The 5$\sigma$ depth of the $g_{475}$ and $I_{814}$ images is 27.5\,mag\,arcsec$^{-2}$ and 27.0\,mag\,arcsec$^{-2}$ respectively. Further discussion of the ACS data is given in \citet{Miley2006}. For our analysis we extract a region of 15"$\times$18.5" around the central radio galaxy which avoids any obvious bright foreground stars. The combined $g_{475}+I_{814}$ image of this region is presented in Fig.~\ref{image}.
The total column density of our galaxy in the direction of the radio galaxy is $4.6\times10^{20}$\,cm$^{-2}$ which results in a galactic reddening of $E(B-V)=0.079$. 

The UV spectral data were obtained using the multi object spectrograph (MOS) on FORS1 in ESO program 64.O-0134(D). These data were obtained as a by-product of the confirmation of the overdensity of \lya\ emitters in the surrounding protocluster \citep{Kurk2000}. The position of the slit on the sky is shown in Fig.~\ref{image}. The spectra were obtained with the 600B grism, and a 1" wide slit, giving a dispersion of 1.2\,\AA\,pixel$^{-1}$ and a pixel scale of 0.2" in the spatial dimension. The resolution of the spectra was approximately 5\AA\,/\,400\kmps. The total exposure time was 6 hours (divided in 6 separate exposures of 1 hour each). The seeing conditions varied between 0.7" to 1.5". The data were reduced using standard IRAF tasks, which included: bias subtraction, flat field removal, and background subtraction. A relative flux calibration was achieved using the spectrophotometric standard star GD108, whilst wavelength calibration was performed using the arc-lamp exposures (He and HgCd lamps). The accuracy of the wavelength calibration is estimated to be 0.2\,\AA.

\subsection{Emission lines within the \gband\ and \iband\ bands}

The \gband\ and \iband\ wavelength ranges both encompass a number of emission lines, notably C{\sc iv}$\lambda$1549 and He{\sc ii}$\lambda$1640 in the \gband, and C{\sc ii}]$\lambda$2326, [Ne{\sc iv}]$\lambda$2424,2426, and Mg{\sc ii}$\lambda$2798 in the \iband. We do not have a spectrum of the emission-line halo surrounding MRC\,1138-262 that covers the lines mentioned redward of He{\sc ii}$\lambda$1640. The C{\sc ii}],  [Ne{\sc iv}] doublet, and Mg{\sc ii} are typically weak: the composite radio galaxy spectrum presented by \citet{McCarthy1993} shows that these lines have equivalent widths of 19\AA, 20\AA\ and 19\AA\ respectively and therefore make up only 7 per cent of the flux within \iband. We therefore assume that the emission line contribution to the \iband\ flux is negligible. \citet{McCarthy1993} give the combined equivalent width of the C{\sc iv} and He{\sc ii} lines of a composite radio galaxy to be 136\AA\ which results in 34 per cent of the flux in the \gband. We therefore use spectra of the \lya\ halo of MRC\,1138-262 to estimate the contribution these lines have to the \gband\ flux.

\begin{figure}
\includegraphics[width=1\columnwidth]{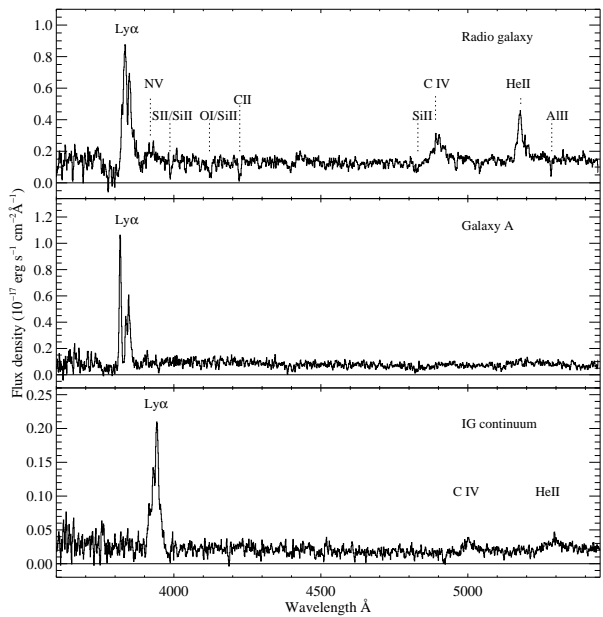}
\caption{ Rest-frame UV spectra of the radio galaxy, a nearby galaxy, and the region within the slit in which low level continuum and/or \lya\ emission was observed. The flux density of the diffuse continuum is per arcsec$^{2}$. The wavelength range of \gband\ is approximately 4100$-$5500\AA, therefore this band encompasses light from the C{\sc iv} and He{\sc ii} emission lines. \label{spectra}}
\end{figure}

To calculate the exact contribution of the emission lines to the \gband\ flux we use the FORS1 long-slit spectrum to examine the UV spectra of 3 distinct regions in the field-of-view: the radio galaxy; a nearby satellite galaxy (labeled galaxy A in Fig.~\ref{image}); and the region in the remainder of the slit (both East and West of the radio galaxy) that did not have any peaked continuum in the spatial direction but had low level continuum and \lya\ emission -- labelled the diffuse continuum region.  The spectra are displayed in Fig.~\ref{spectra} and show that the emission line contribution to the \gband\ flux is not uniform over the entire field-of-view. The equivalent width of the emission lines in all three regions are given in Table \ref{ew}. The central radio galaxy exhibits the emission lines with the largest equivalent width; N{\sc v}, C{\sc iv} and He{\sc ii} are prominent. Although galaxy A has a large \lya\ equivalent width it does not exhibit any metal emission lines, therefore emission lines would not contribute to the flux in the \gband\ passband. The diffuse continuum region shows strong \lya\ emission as expected since a large \lya\ halo encompasses the radio galaxy. Both C{\sc iv} and He{\sc ii} are detected with a combined equivalent width of 71\AA, although there are variations in the strength of these lines along the slit length. As we are primarily interested in the nature of the intergalactic light (IGL) we adopt 71\AA\ as the emission line contribution which means the emission lines are responsible for 17 per cent of the total \gband\ flux. This contribution is  subtracted from the \gband\ flux in the rest of this article. Since the contribution of the metal lines to the Spiderweb system halo is almost half that of the \citet{McCarthy1993} composite radio galaxy spectrum, we assume the contribution of metal lines within the \iband\ filter to be negligible. The spectrum of the radio galaxy shown in the top panel of Fig.~\ref{spectra} shows numerous absorption lines redward of \lya. Because the interstellar absorption lines are generally stronger than stellar absorption lines, and none of the lines from the radio galaxy show P-Cygni like profiles, we assume that these lines are interstellar and identify them as S{\sc iii}/Si{\sc ii}$\lambda1260$, O{\sc i}/Si{\sc ii}$\lambda1303$, C{\sc ii}$\lambda1335$, Si{\sc ii}$\lambda1527$ and Al$\lambda1671$ originating from gas at $z=2.164$.

\begin{table}
 \centering 
\begin{tabular}{l|ccccc}
  \hline
  Region & size (arcsec$^{2}$)&Ly$\alpha$ &N{\sc V}&C{\sc iv}&He{\sc ii}\\  \hline 
 Radio galaxy&2.4& 147&21&57&44\\
 Galaxy A& 2.4&124&--&--&--\\
 Diffuse Continuum& 7.2&202&--&35&41\\
\hline
\end{tabular}
\caption{Equivalent width of rest-frame UV emission lines in \AA\ from three separate regions: the radio galaxy, a nearby galaxy and the region in the slit in which low level continuum and/or \lya\ emission were observed.  The sizes of the extracted regions are given in the second column. \label{ew}}
\end{table}

\subsection{Extraction of galaxies}
\label{extract}
\citet{Pentericci1998} concluded that the UV bright clumps surrounding the radio galaxy were star forming galaxies. Defining which pixels belong mainly to a galaxy in this region is rather arbitrary since the bright `clumps' do not have regular shapes. We used {\sc Sextractor} \citep{Bertin1996} in the double image mode, where we use a detection image, constructed as the inverse variance weighted average of the $g_{475}$ and $I_{814}$ images, to determine the apertures, whilst photometry is carried out on the  individual filter images. A map of the total exposure time per pixel was used as the detection weight map. Regions with 5 or more connecting pixels which had a S/N greater than 3 times the local background RMS standard deviation were extracted. The right panel of Fig.~\ref{image} shows the 15"$\times$18.5" region surrounding the radio galaxy with all regions identified as 'galaxies' by {\sc Sextractor} removed and colored white. AB Magnitudes of the galaxies were determined from 
\begin{eqnarray}
I_{814}= -2.5 \log(counts) + 36.841-0.074\\
g_{475}= -2.5 \log(counts) + 36.856-0.15, 
\end{eqnarray}
\citep{Sirianni2005}. The last numbers on the right-hand side of both equations corrects for Galactic extinction.
The slope of the UV continuum is defined as $\beta$ given by $f_\lambda=\lambda^{\beta}$. Therefore a spectrum that is flat in $f_\nu$ has  $\beta=-2$. We calculate the UV slope $\beta$ through
\begin{eqnarray}
\beta=\frac{-0.4\times(g_{475}-I_{814})}{\log_{10}\frac{\lambda_{475}}{\lambda_{814}}}-2,
\label{beta}
\end{eqnarray}
where $\lambda_{475}$ and $\lambda_{814}$ are the effective bandpass wavelengths.
Errors were calculated using the RMS images from {\it Apsis} and are quoted to 1$\sigma$ unless noted otherwise.

\begin{figure}
{\includegraphics[width=1.0\columnwidth]{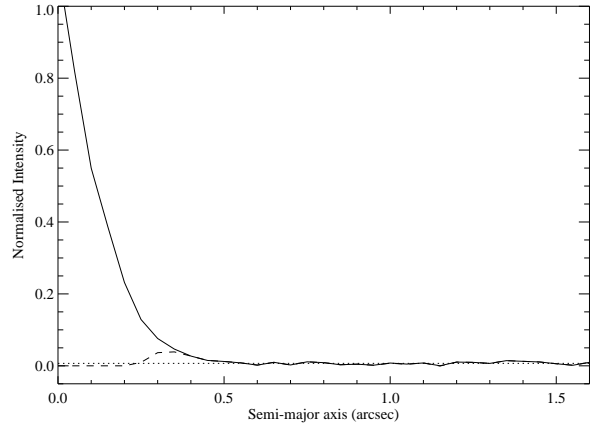}}
\caption{\iband\ radial surface brightness profile of the galaxy that lies at position (2.3,\,6.5) in Figs.~\ref{sb} and \ref{color}. Solid line is the radial profile of the galaxy, whereas the dashed line is the radial profile of the same region with the galaxy removed by {\sc Sextractor}. The dotted line indicates the continuum and background level. The area between the dashed and the dotted line is a measure of the galaxy flux that is not removed.\label{gal2_profile}}
\end{figure}

Since the galaxies were detected and extracted using {\sc Sextractor} isophotal apertures, only those pixels above a certain flux threshold were removed. Some of the flux in the wings of the galaxy profile is not removed by this process. As many of the galaxies have irregular profiles we cannot simply fit an ellipse to the galaxies and remove all flux within the ellipse. Instead we estimate how much galaxy flux typically remains after the detected object is removed by {\sc Sextractor} using three galaxies which have the most elliptical profiles compared to the other galaxies in the Spiderweb system. Azimuthally averaged radial profiles of the three galaxies were created, and compared to profiles of the same region after the galaxies were removed using our {\sc Sextractor}-based algorithm. Fig.~\ref{gal2_profile} shows an example of the radial profile of one galaxy (solid line) compared to the radial profile of the same region but with the galaxy masked out (dashed line). The dotted line indicates the continuum and background level. Fig.~\ref{gal2_profile} clearly shows that some galaxy flux is not removed. The galaxy flux remaining is estimated as the area under the dashed profile. On average the remaining galaxy light totals 27\,mag\,arcsec$^{-2}$ in \iband\ and  29\,mag\,arcsec$^{-2}$ in \gband. The galaxy's wings extend beyond the radius of the galaxy as defined by {\sc Sextractor} across an area of  approximately 0.6\,arcsec$^{2}$. Since the surface brightness of the galaxy's wings in \gband\ is 1.5 magnitudes below our definition of the detection limit of the intergalactic light (see section \ref{voronoi}), this remaining galaxy flux will not contribute greatly to the intergalactic light. However, the flux in the wings of the galaxies is brighter in \iband\ compared to \gband, therefore the \gband--\iband\ colour can be affected: the regions surrounding the galaxies would appear to be redder due to the presence of the galaxy wings. No correction has been made for this effect, but the possibility of associated flux from the galaxies wings should be kept in mind when considering Figs.~\ref{sb} and \ref{color}.

\subsection{Voronoi 2D binning}
\label{voronoi}
To increase the signal-to-noise (S/N) the images were re-binned using the \gband\ image. The galaxies and the intergalactic regions were binned separately. The pixels of the galaxies were binned using the Voronoi 2D-binning method by \citet{Cappellari2003} such that each bin had a S/N of 12.  Many of the pixels at the centre of the galaxies had a S/N much greater than 12, therefore they were not binned together with another pixel. The intergalactic region had a lower S/N per pixel so the pixels were binned such that each bin had a S/N of 5 using the Voronoi 2D-binning method by \citet{Cappellari2003} with the Weighted Voronoi Tessellation modification proposed by \citet{Diehl2006} instead of the Centroidal Voronoi Tessellation. We checked that there was no difference in any of the results if we binned the data using the $I_{814}$ image. Using a much higher S/N (e.g.\,10) resulted in very large bins which meant we lost a great deal of positional information.

Since we wish to observe the intergalactic UV light we enforce that UV diffuse emission is detected in a bin only if the bin has a surface brightness brighter than 27\,mag\,arcsec$^{-2}$ in \iband and 27.5\,mag\,arcsec$^{-2}$ in \gband. This corresponds to a firm 5$\sigma$ detection of the IGL in the rest frame near-UV (NUV; 2215$-$3000\AA) and far-UV (FUV; 1300--1740\AA) bands above 5$\sigma$ confidence. In the rest of the article we refer to the IGL as the region in which the \iband\ flux is brighter than 27\,mag\,arcsec$^{-2}$ and the \gband\ flux is brighter than 27.5\,mag\,arcsec$^{-2}$, but not within a galaxy.

\section{Results}
\label{results}
\subsection{Distribution of the UV light}

\begin{figure*}
{\hspace{-1in}\includegraphics[width=1.75\columnwidth]{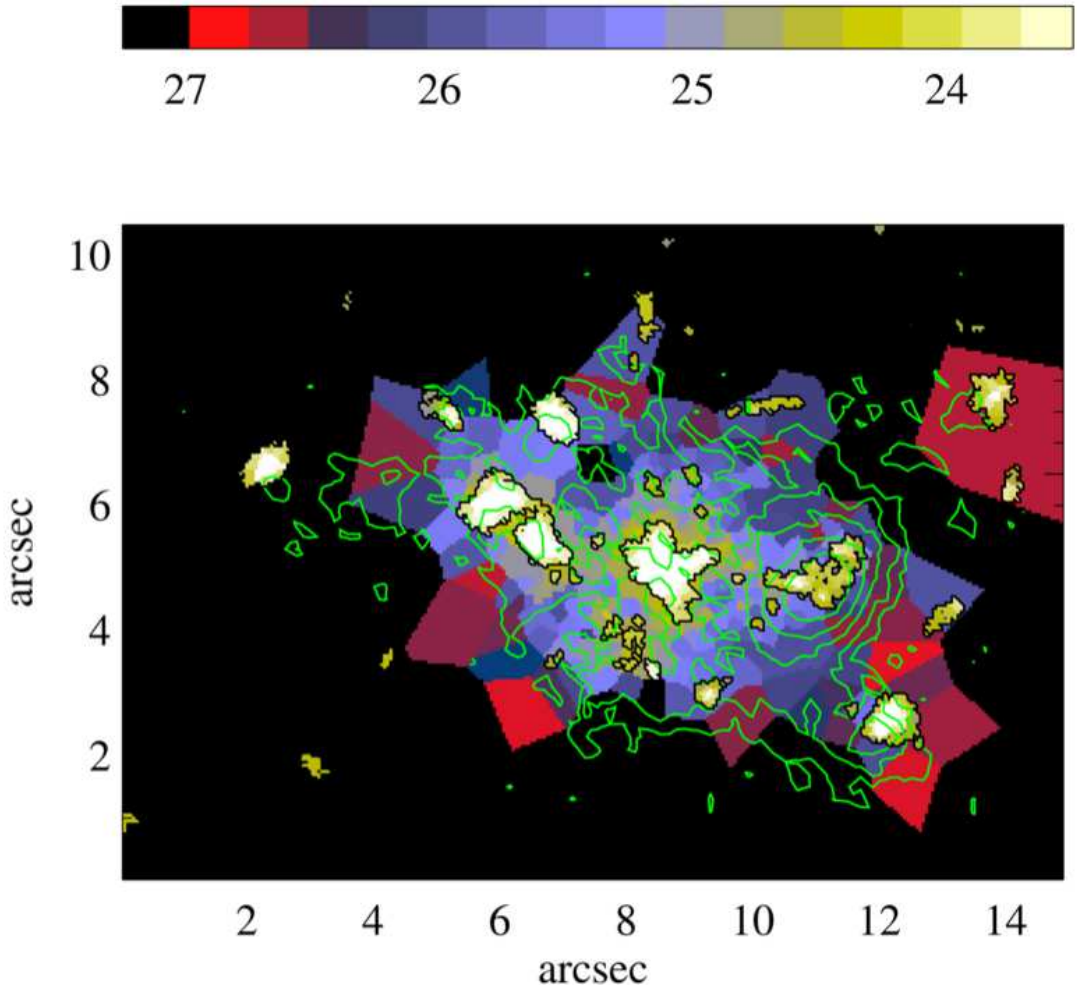}}
\caption{\iband\ surface brightness of the Spiderweb system. Colour bar scale is in mag\,arcsec$^{-2}$. Data has been binned in 2D so that each bin within a galaxy has a S/N greater than 10, and each bin outside a galaxy has a S/N of $\sim$5 as described in section \ref{voronoi}. Galaxies are outlined in black. Green contours are \lya. The \lya\ flux peaks on the galaxy that lies $\sim$3\, arcsec West of the radio galaxy. Image is rotated 5 degrees East from North, with North Up and East left.\label{sb}}
\end{figure*}

{\sc Sextractor} identifies 48 separate objects as possible galaxies, although only 25 are detected above the RMS noise. As we are interested in the light in between the galaxies we remove all 48 detections of structure. The data in both bands were binned to a uniform S/N as described in section \ref{voronoi} and the surface brightness of the galaxies and the IGL are presented in Fig.~\ref{sb}. We define a detection of  IGL as a bin which has 5$\sigma$ above the RMS background level in both \gband\ and \iband. This corresponds to 27\,mag\,arcsec$^{-2}$ in \iband\ and 27.5\,mag\,arcsec$^{-2}$ in \gband. Because of this strong limit there may exist a fainter component that we do not have the depth in the images to unambiguously detect. In the \iband\ there appears to be residual light around some of the galaxies therefore the stellar halos of these galaxies extend further than extracted by the {\sc Sextractor} algorithm (see section \ref{extract}).  Black areas in Fig.~\ref{sb} are those in which the surface brightness is less than 27\,mag\,arcsec$^{-2}$ in \iband\ or 27.5\,mag\,arcsec$^{-2}$ in \gband. The IGL surrounds the central radio galaxy in an ellipsoid shape, with best fit ratio 0.62 and angle 148.9$^{\circ}$, centered on the core of the radio galaxy and extending over 60\,arcsec$^{2}$ (4243\,kpc$^{2}$). Removing the galaxies within this region results in a total IGL area of 52\,arcsec$^{2}$ (3168\,kpc$^{2}$) The orientation of the intergalactic UV light is both similar to the \lya\ halo and within 30$^\circ$ of the radio axis. There is some extended light surrounding two other galaxies (one to the North, the other in the South,  neither shown in Fig.~\ref{sb}), but generally the IGL is limited to the neighborhood of the radio galaxy and the group of central satellite galaxies.  

The total amount of continuum in the IGL is  21.7$\pm0.1$\,mag in \iband\ and 22.4$\pm0.15$\,mag in \gband. 39$\pm7$\% of the total \iband-band light exists as the IGL, whilst 44$\pm12$\% of the total \gband-band light exists in the intergalactic region. 

\subsection{Intergalactic UV colour}
\begin{figure*}
{\hspace{-1in}\includegraphics[width=1.75\columnwidth]{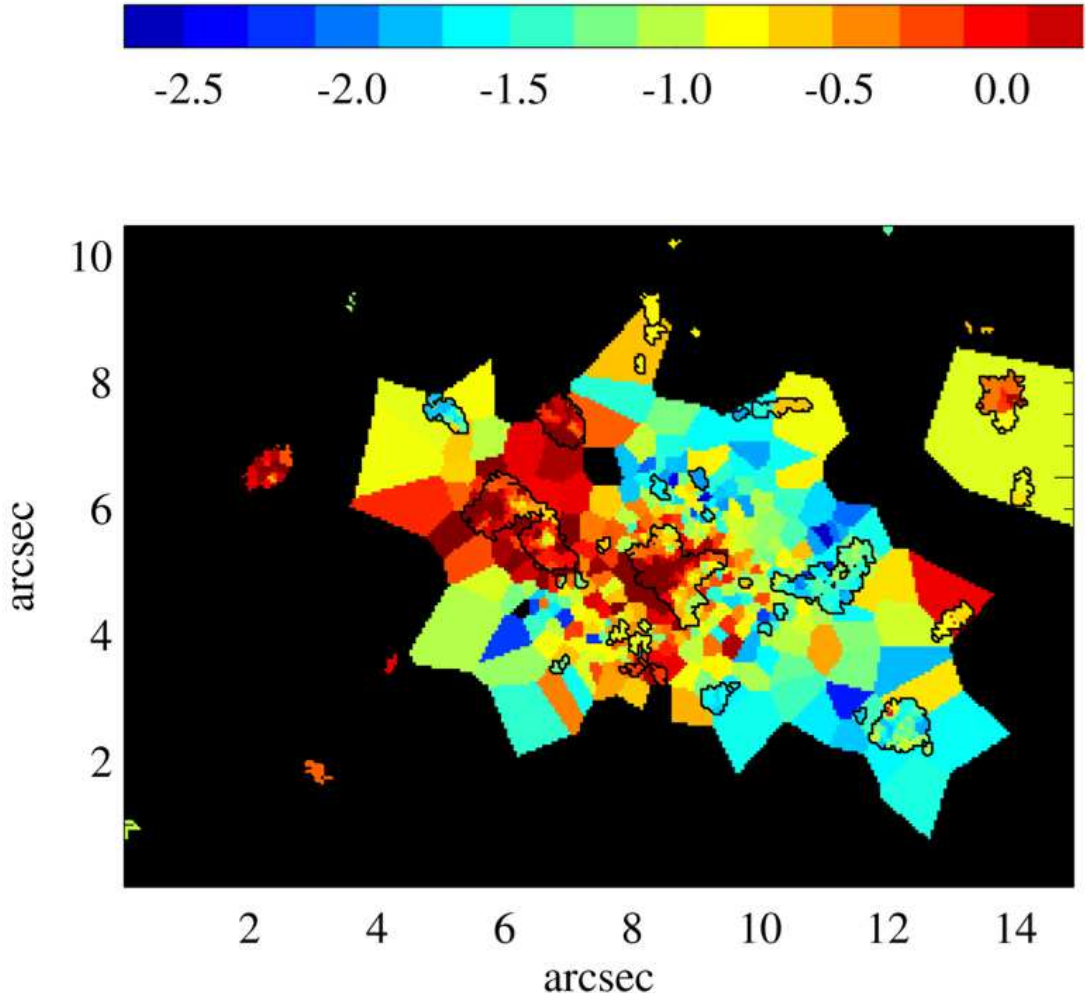}}
\caption{Slope of UV spectrum ($\beta$) of galaxies and intergalactic light of the Spiderweb system.  Data has been binned in 2D as described in section \ref{voronoi}. Galaxies are outlined in black. The \lya\ flux peaks on the galaxy that lies $\sim$3\, arcsec West of the radio galaxy which has the bluest colour compared to the other satellite galaxies. Image is rotated 5 degrees East from North, with North Up and East left. \label{color}}
\end{figure*}
Fig.~\ref{color} plots the UV slope $\beta$ for all bins which have an \iband\ surface brightness greater than 27\,mag\,arcsec$^{-2}$ and a  \gband\ surface brightness brighter than 27.5\,mag\,arcsec$^{-2}$.  It is difficult to distinguish the galaxies from the IGL by their UV slope alone: the \iband-\gband\ colour of the galaxy and the {\it immediate} area surrounding each galaxy are similar. For example the three red satellite galaxies to the East and Northeast of the central radio galaxy are surrounded by red IGL, whereas the blue galaxy to the West of the radio galaxy is surrounded by blue IGL.

\begin{figure}
\includegraphics[width=1\columnwidth]{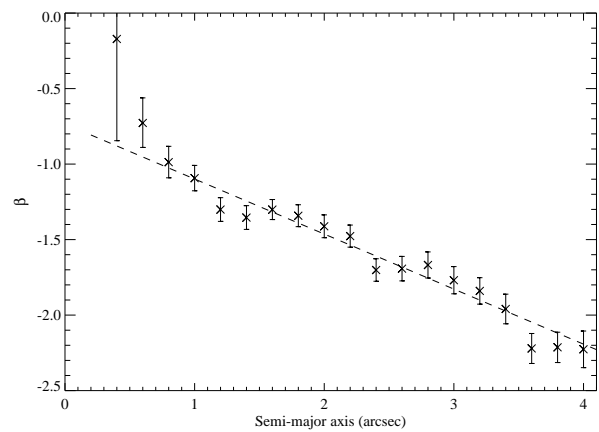}
\caption{Azimuthally averaged radial profile of the slope of the UV spectrum $\beta$. Errors are 3$\sigma$. The solid line has a slope of -0.37 and is the error-weighted best fitting line to the data points. The IGL becomes bluer toward larger radii. \label{color_profile}}
\end{figure}

The average slope of the UV spectrum of the IGL is $\beta_{gI}=-0.84\pm0.5$, but varies greatly from region-to-region. Fig.~\ref{color_profile} plots the azimuthally averaged radial profile of the intergalactic UV light (all galaxies identified by {\sc Sextractor} have been removed). It clearly shows that the colour of the IGL becomes bluer with increasing distance from the radio galaxy (marked by a white cross in Fig.~\ref{image}). 

\begin{figure}
\includegraphics[width=1\columnwidth]{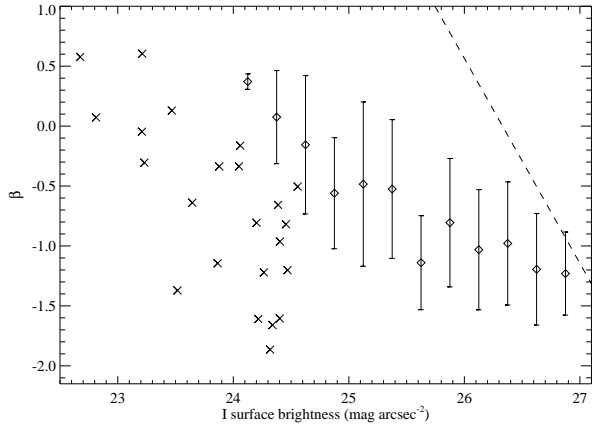}
\caption{Crosses are galaxies, diamonds are IGL surface brightness grouped  in bins of 0.25\,mag\,arcsec$^{-2}$. The dashed line shows the detection limit imposed by the 27.5\,mag depth limit in the \gband. The slope of the best fitting line to the galaxies is $-0.9\pm0.2$ and the correlation co-efficient is $-0.76$. The best fitting line to the azimuthally averaged intergalactic light has a slope of $-0.7\pm0.1$. The slopes of both fits agree within 1$\sigma$ of the error of the line fitting. Error bars on diamond points indicate 1$\sigma$ of the distribution in each surface brightness bin. \label{gal_slope}}
\end{figure}

Fig.~\ref{gal_slope} plots the UV continuum slope $\beta$ against the \iband\ surface brightness. The crosses are the galaxies, and the diamond symbols are the averaged IGL grouped in bins of 0.25\,mag\,arcsec$^{-2}$. The galaxies show a clear $\beta$--surface brightness relation with a correlation co-efficient of $-0.76$. The $\beta$-magnitude relation (not plotted) also shows a correlation but with a less significant co-efficient of -0.43. A similar relation is seen in a number of works \cite[e.g][]{Meurer1999,Ouchi2004}, and possible causes may be a mass-metallicity, mass-extinction, or mass-age relation. This figure shows that the galaxies are generally redder than the IGL. The average slope of the UV spectrum is $\beta_{GI}=-0.84\pm0.5$, whereas 60\% of the galaxies in the Spiderweb system have a redder colour than this. Whilst the average colour of the total IGL is bluer than that of the average galaxy colour, the IGL directly surrounding a galaxy is often similar in colour to the galaxy itself. Therefore the IGL furthest away from a galaxy must have a bluer colour than the average galaxy colour.

\section{The origin of the intergalactic light}
\label{nature}
We have established the existence of low surface brightness UV light situated in between the galaxies of the Spiderweb system. Determining the origin of this IGL has parallels to the problem of the UV continuum excess of high redshift radio galaxies (see \citealt{Tadhunter2002} and references therein). Numerous mechanisms can contributed to the UV excess including: scattered AGN light, young stellar populations, and nebular continuum emission. And the relative contribution of each mechanism varies between radio sources \citep{Solorzano2004}. We now determine the origin of the IGL by investigating the possible emission mechanisms.
\subsection{Faint cluster galaxies}
\begin{figure}
\includegraphics[width=1\columnwidth]{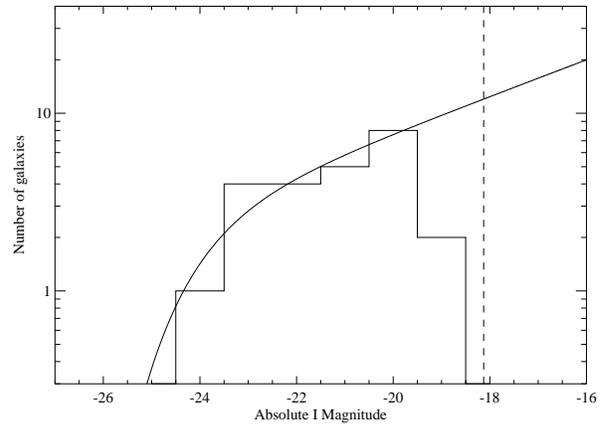}
\caption{The luminosity function of the galaxies. The solid line is a Schechter function fit to the data (see text for details). The dotted line indicates how much of the luminosity function is missing if the intergalactic light results from a population of fainter galaxies; approximately 15 galaxies with absolute I magnitude (rest frame NUV) between $-19.5$\,mag and $-18.1$\,mag would provide the necessary light. \label{lum_func} }
\end{figure}

We examine the possibility that the intergalactic UV light may be caused by a fainter population of cluster galaxies that lie between the detected satellite galaxies. Fig.~\ref{lum_func} plots the rest-frame NUV (\iband) luminosity function of the detected galaxies with the solid line showing the best fitting Schechter function $\phi(M)=0.4\phi^{*}{\rm log}_{10}[10^{0.4(M^{*}-M)}]^{\alpha-1}\exp[-10^{0.4(M^{*}-M)}]$ \citep{Schechter1976}. The best fitting parameters were found to be $M^{*}=-24.3$,  $\phi^{*}=198.3$\,mag$^{-1}$Mpc$^{-3}$ and the slope $\alpha= -1.26$.  The slope of the faint end, $\alpha$,  is found to be similar to that of a sample of $z=2.2$ galaxies from \citet{Sawicki2006}. The bright $M^{*}$ and large $\phi^{*}$ are a direct consequence of the fact that the region of interest is associated with a massive galaxy at the centre of a forming cluster, and should not be taken as a measure of the luminosity function beyond this region. The luminosity function is complete to --19.5\,mag and encompasses almost all of the galaxies detected, therefore the integral of the Schechter function from $-19.5$ to $-24.5$ results in the total \iband\ flux from the galaxies in the Spiderweb complex. The intergalactic UV light in this bandpass accounts for 39\% of the total flux above a surface brightness of 27.0\,mag\,arcsec$^{-2}$. Therefore if the IGL resulted from a population of fainter cluster galaxies, the missing galaxies would fill up the Schechter function up to $-18.1$\,mag (indicated by the dotted line in Fig.~\ref{lum_func}). Therefore the IGL would comprise of approximately 15 galaxies with absolute magnitudes between $-19.5$\,mag and $-18.1$\,mag.  The IGL extends across a region 4243\,kpc$^{2}$ ($\sim60$\,arcsec$^2$). Therefore each of the 15 faint galaxies would have to extend across $\sim$4\,arcsec$^{2}$ --  larger than the massive radio galaxy, or be distributed in a very ordered arrangement around the radio galaxy. Such an ordered distribution is unlikely and any faint galaxies are not likely to have such a large extent, therefore the IGL cannot be caused by a non-detected faint cluster galaxy population, but rather it must be due to an intrinsically diffuse source or very many unresolved radiation sources.

\subsection{Nebular continuum emission}
The distribution of the UV continuum and the \lya\ halo are similar which suggests that the continuum light may be produced by nebular continuum. Nebular continuum emission results from a combination of two-photon continuum emission, free-free emission and free-bound recombination continuum, as well as a pseudo-continuum created by blended higher order Balmer lines just above the 3646\AA\ jump. Theoretical case B nebular continuum emission  can be calculated using tabulated emission coefficients \cite[e.g][]{Brown1970} and the measurement of the H$\beta$ emission, therefore it can be well-constrained. We do not have a deep H$\beta$ measurement of the \lya\ halo hence we estimate the average H$\beta$ flux across the region from the measured He{\sc ii}$\lambda1640$ flux in the diffuse continuum spectrum (Fig.~\ref{spectra}). We use the {\sc starlink~dipso} command {\sc nebcont} to estimate the contribution of the nebular continuum. Assuming a gas temperature of 15,000\,K, density of 100\,cm$^{-3}$ and 10\% He abundance, the  He{\sc ii} flux of $8.46\times10^{-18}$ \ergpcmsqpspsqarcsec\ results in a nebular continuum in \gband\ of $\sim$31.0\,mag\,arcsec$^{-2}$ or $\sim$30.8\,mag\,arcsec$^{-2}$  in \iband. Therefore less than 3\% of the intergalactic UV light in both bands result from nebular continuum. 

\subsection{Scattered light}
Previous studies have suggested that the UV excess in radio galaxies may arise from scattered quasar light \cite[e.g][]{Fabian1989} and early polarimetric  observations showed high degrees of linear polarization agreeing with the quasar scattering theory \citep{Tadhunter1992,Cimatti1993,Vernet2001}. Detailed studies of dimmer radio galaxies show that the contribution from scattered quasar light can be relatively minor \citep{Tadhunter2002}. Here we examine the possibility that the diffuse IGL may result from scattered light originating from either a hidden quasar or the stellar population of the radio galaxy  as well as the numerous satellite galaxies. We first discuss whether it is energetically possible for the IGL to result from either electron scattering or dust scattering of star light or a hidden quasar.

The optical depth, $\tau$, of electron scattering is 
\begin{equation}
\tau=\sigma_{\rm T}{\rm Rn}_{\rm e},
\label{tau}
\end{equation}
 where $\sigma_{\rm T}$ is the Thompson cross section (6.7$\times10^{-25}$\,cm$^{2}$), R is the radial distance and $n_{\rm e}$ is the electron density. The density of the gas in the \lya\ halo is estimated to be $\sim$100\,cm$^{-3}$ but the volume filling factor is $f_{v}\sim10^{-5}$ \citep{Villar-Martin2003}. Therefore the average density over the halo volume is only $10^{-4}-10^{-3}$\,cm$^{-3}$. However, any surrounding hot intergalactic medium ($\sim1\times10^{7}$\,K) may have a higher density ($\sim0.15$\,cm$^{-3}$) if we assume that it  is in pressure equilibrium with the \lya\ emitting material. Since we wish to determine whether electron scattering is possible we use the highest plausible value of ${\rm n_{e}}$ in equation \ref{tau}, i.e.~that of the hot intergalactic medium. The light must be scattered up to 25\,kpc from the central radio galaxy, therefore $\tau \sim 7.8\times 10^{-3}$. Assuming all the scattered light is observed as the IGL allows us to place a lower limit to the source intensity of 15.7\,mag through
\begin{equation}
I_{\rm source}=\frac{\rm IGL}{(1-e^{-\tau})}~~.
\end{equation}
 The sum of all the UV light emitted from the galaxies (radio galaxy and the satellites) is only \iband=21.2\,mag, therefore the IGL cannot be produced by electron scattering of the stellar light from the radio and satellite galaxies. Bright quasars at a redshift of 2 typically have an apparent magnitude of $\sim19$, and the brightest is $\sim$16\,mag \citep{Schneider2007}. Therefore it is energetically possible, although unlikely, that the IGL is produced by electron scattering of the light from a very bright hidden quasar. 

Dust has a far greater scattering efficiency than electrons, therefore if any dust exists in the halo, it will dominate the scattering process. Homogeneous optically thin dust scattering predicts a blueing of the emerging spectrum due to the wavelength dependance of the scattering efficiency of small dust grains. The IGL is observed to become progressively bluer with distance from the central radio galaxy (Fig.~\ref{gal_slope}) which may arise from multiple scattering. Models from \citet{Cimatti1993} of dust-scattering of quasar UV light, where the scattered light is approximately proportional to the galactic light, indicate a dust mass of $\sim10^8$\Msun\ is required to scatter enough light. \citet{Vernet2001} predict that only 7\% of the incident light is scattered by $10^8$\Msun\ of dust in a spherical distribution of radius 40\,kpc from the central source. Therefore a quasar of luminosity of $3\times10^{31}$\ergpsphz\ (apparent magnitude of 19 at $z=2.156$)  is required to illuminate the halo. It is plausible that MRC\,1138-262 contains a quasar of apparent magnitude 19 and submillimetre observations of both  Stevens et al.~(2003) and \citet{Greve2006} show that MRC1138-262 contains spatial extended dust of total mass between 2 and $4.6\times10^8$\Msun. Therefore it is energetically possible that dust scattering of a quasar produces the IGL. Assuming a dust-to-gas ratio applicable to our Galaxy, this dust mass predicts a gas mass of $10^{11}$\Msun, which is more than an order of magnitude greater than the mass of ionized gas measured in this system: $2.3-6.5\times 10^{9}$\Msun \citep{Nesvadba2006}). Therefore a cooler gas component must also exist in the halo.  Evidence for cold gas is observed as absorption in the \lya\ emission lines of the radio galaxy, satellite galaxies and IGL shown in the spectra of Fig.~\ref{spectra}.

We now consider whether the model of dust scattering of quasar light is consistent with the morphology and colour of the IGL. A broad luminous \ha\ emission line with a full width at half maximum of $14,900$\kmps\ is a prominent feature in the AGN spectrum of MRC\,1138-262 \citep{Nesvadba2006}, however, the rest-frame UV spectrum (Fig.~\ref{spectra}) reveals no nuclear broad emission lines. Additionally, \citet{Nesvadba2006} measure a large extinction ($A_V=8.7$) from the luminous broad Balmer emission lines. This indicates that MRC\,1138-262 harbours an obscured quasar. Therefore we expect any scattered light to be emitted from a biconical structure, revealing a butterfly morphology such as that observed in a number of low redshift Seyfert galaxies including the nearby NGC\,5252 \citep{Tsvetanov1996}. The smooth elliptical morphology of the MRC\,1138-262 IGL is not consistent with bi-conical scattering cones, therefore the morphology of the IGL argues against quasar scattered light as a possible source for the light. In the clumpy torus  model \citep{Elitzur2004}, where the torus is made of many small dense clouds, there is a finite probability that an unobscured line-of-sight to the AGN can occur through the torus. There have been a few recorded transits between type 1 and type 2 AGN (e.g.~\citealt{Storchi-Bergmann1993}) that may result from such an opening in the torus. If such a situation occurred in MRC\,1138-262 the morphology of the scattered light would be elliptical \citep{Villar-Martin2007} as seen in the Spiderweb system. This scenario is unlikely as transitions from type 1 to type 2 are rare, however this remains a possible explanation for the IGL.

A further argument against the IGL resulting from quasar scattered light is provided by the uneven variation of the colour of the IGL.  The IGL immediately surrounding a galaxy is very similar in colour to the galaxy (even though on average the IGL is bluer than the average galaxy colour). Such a clear correlation between the colour of the IGL and spatial position of the galaxies would require an unpleasant amount of fine-tuning if the IGL originates from quasar scattered light. Therefore the morphology and colour distribution of the IGL suggests that it does not originate from scattered quasar light. 

To summarise: the IGL cannot be produced by scattering (by either electrons or dust) of the stellar light of the radio and satellite galaxies as the scattering efficiency is too low. Scattering of quasar light is ruled out because MRC\,1138-262 is believed to harbour a hidden quasar and the elliptical morphology of the IGL is not consistent with scattering from a hidden quasar. Furthermore the colour of the IGL shows a correlation with the colour of the satellite galaxies, which is difficult to explain if the IGL resulted from scattered quasar light. We therefore conclude that the IGL does not result from scattered light from either a hidden quasar or the stellar light of the galaxy population.

Scattering would result in high polarization, therefore imaging polarimetry measurements of the Spiderweb system would determine the importance and extent of any quasar scattered light.

\subsection{Synchrotron, inverse Compton, and synchrotron self-Compton emission}
\subsubsection{Synchrotron emission}
The radio emission and spectral index can be used to determine whether the UV continuum emission may arise from the extrapolation of the synchrotron radio emission to UV wavelengths. The flux at 5\,GHz is 1.6$\times10^{-24}$\ergpspcmsqphz\ \citep{Carilli1997} and the integrated spectral index is measured to be $-1.8$ from the 4.2GHz to 8.1GHz range \citep{Pentericci1997}. This value is corroborated by \citet{Carilli1997}. The core has a flatter spectral index of $-1.2$ compared to the hotspots and extended emission. Extrapolating to 2600\AA, corresponding to the central wavelength of the \iband, results in a \iband\ flux of 35.0\,mag. Additionally this mechanism could not explain the IGL beyond the observed radio structure. Therefore the UV continuum cannot be synchrotron emission.
\subsubsection{Inverse Compton}
Another possibility is the up-scattering of the cosmic microwave background photons (CMB) by the relativistic electrons of the radio source. The energy density of the CMB increases at higher redshift by a factor of $(1+z)^4$, therefore inverse Compton emission may be significantly more important at higher redshift. The scattering electrons are presumed to belong to the same distribution as those producing the synchrotron radio emission. Therefore the amount of UV flux $S_{UV}$, at frequency $\nu_{UV}$, that can be produced through inverse Compton can be estimated through eqn.~11 of \citet{Harris1979} :
\begin{equation}
S_{UV}=\frac{(5.05\times10^4)^\alpha C(\alpha)G(\alpha) (1+z)^{3+\alpha}S_R\nu_R^{\alpha}}{10^{47}B^{1+\alpha}\nu_{UV}^{\alpha}}~~~~~=21.7\,{\rm mag}, 
\label{uv_flux}
\end{equation}
where $\alpha$ is the absolute value of the spectral index (1.8), $C(\alpha)$ is a constant approximated to be 1.15$\times10^{31}$, $G(\alpha)$ is a slowly varying function of $\alpha$ ($G=1.08$ for $\alpha=1.8$), $S_R$ is the radio flux at frequency $\nu_R$, and $B$ is the magnetic field  strength. The minimum energy magnetic field strength $B_{eq}$ has been estimated though equipartition arguments to be $\sim200\mu$G \citep{Carilli1998}. 
Therefore it is energetically plausible that the UV extended emission can be produced by inverse Compton emission.  It should be noted that the expected flux would be several orders of magnitude fainter if the spectral index is flatter. This up-scatting of the photons would require electrons with $\gamma\sim1000$. Electrons with $\gamma=1000$  would quickly lose energy through inverse Compton and synchrotron radiation in the environment of the IGL. It is difficult to devise a scenario where such a large volume is continuously populated by a diffuse medium of high-energy electrons.

Although inverse Compton emission is energetically plausible, both the morphology and colour of the IGL rule out inverse Compton as a plausible mechanism.  The morphology of the IGL does not match the spatial distribution of the radio emission: the radio emission comes from the core, hotspots and jet-like regions, whereas the UV IGL has an ellipsoid shape centered on the radio galaxy and is brighter along lines connecting the satellite galaxies. The radio morphology changes very little between 8.2GHz, 4.7GHz, and 1.4GHz, and we assume that the structure of the radio source at low frequencies remains the same as that at high frequencies. Therefore we currently rule out inverse Compton emission based on morphological arguments, however, if a low frequency radio halo is discovered with the new low frequency radio observatories e.g.~LOFAR, we must reconsider this conclusion. 

The most solid argument against inverse Compton emission relies on the slope of the UV spectrum of the IGL and its relation to the electron distribution index $P$ defined as\begin{equation}
N(E)dE=CE^{-P}dE~~~~~~E_1<E<E_2
\end{equation}
where $N(E)$ are the number of electrons with energy $E$ between $E_1$ and $E_2$. The spectral index ($S$) of the emission produced by single scattering of inverse Compton is related to the electron distribution index $P$ by
\begin{equation}
S=\frac{P-1}{2}=\beta+2,
\end{equation}
\citep{Rybicki1979}, and is identical to the relationship between the spectral index ($\alpha$) of the Synchrotron emission produced by the same population of electrons with distribution index $P$. $S$ is trivially related to the observed slope of the UV spectrum, $\beta$. The observed range of $\beta$ lies between $-2$ and $1$, which corresponds to $0<S<3$ or $1<P<7$. Large variations in the colour of the IGL occur across distances as small as a few kpc and such a large intrinsic range in $P$ across these distances seem unlikely. Furthermore given the strong relation between $\alpha$ (or $P$) and the flux emitted in the UV (eqn.~\ref{uv_flux}) any electron population which has $P<3$ would contribute negligibly to the UV flux. Therefore regions of the IGL with $\beta<-1$ cannot be produced by inverse Compton emission. Alternatively we can imagine a situation where the IGL is produced by a population of electrons described by a single $P$ and the range in $\beta$ produced from reddening by intrinsic dust. As a consequence of this we should expect to find a correlation between the UV flux and slope of the UV spectrum, such that the brighter regions would be bluer, but Fig.~\ref{gal_slope} shows the reverse correlation with the brighter regions having a redder colour.  Therefore the colour of the IGL rules out inverse Compton emission as a possible origin of the IGL.
\subsubsection{Synchrotron self-Compton emission}
Synchrotron self-Compton emission occurs when the radio photons are up-scattered by the same population of electrons that produced the synchrotron radio emission. Although synchrotron self-Compton emission can also be ruled out on morphological grounds, we examine whether it is energetically possible. A condition for synchrotron self-compton to occur is that the ratio of the energy loss of an ultrarelativistic electron by inverse Compton to the synchrotron radiation ($\eta$) must be greater than 1, i.e
\begin{equation}
\eta=\frac{U_{rad}}{B^{2}/2\mu_{0}}=\frac{L_{Tot}}{4\pi r^{2}c} \frac{2\mu_{0}}{B^{2}}> 1.
\end{equation}
$U_{rad}$ is the radiation energy density, $B$ is the magnetic field strength measured in this hotspot through equipartion ($10^{-3}$\,Gauss, \citealt{Pentericci1997}),  $\mu_{0}$  is the permeability of free space, $L_{Tot}$ is the total radio luminosity in the hotspot, and $r$ is the size of the clump. We use the parameters of the UV continuum clump that lies $\sim$3\,arcsec West of the radio galaxy and matches the position of a radio hotspot. The radio luminosity $L_{Tot}=9\times10^{44}$\ergps \citep{Pentericci1997} and $r$ is  estimated from the UV emission to be $\sim1\arcsec^{2}\sim$70\,kpc$^{2}$. This results in $\eta<1\times10^{-4}$, therefore is not possible that the radio emission produced by the relativistic photons are scattered to UV energies by the same population of electrons.
\subsection{Intergalactic star formation}
The final possibility is that the IGL results from UV emission from hot young stars. There are two possibilities for the origin of these stars: either they were stripped from the star forming satellite galaxies, or the star formation occurred in-situ in the halo. The high rest-frame UV continuum suggests that we are observing OB stars, which have a main sequence lifetime of $\sim10$\,Myrs and are thus good tracers of recent star formation. Stars recently  stripped from galaxies  would have a similar colour to that of the host galaxies, however the stripped stars would age passively whilst the galaxies continue to form stars, therefore the IGL should be redder than the galaxies in general if the IGL results from stripped stars. Since OB stars produce the vast majority of the UV emission and yet have a very brief lifetime, the colour of the UV light rapidly becomes redder once the star leaves the star forming region. Although the colours of the IGL immediately surrounding a galaxy is very similar to the colour of the galaxy itself, generally the IGL is bluer than the galaxies as shown  in Fig.~\ref{gal_slope}. The virial velocity of a $10^{12}$\Msun\ galaxy, assuming a radius of 10\,kpc is 650\kmps.  An OB stripped star with a typical velocity of $\sim650$\kmps\ would travel a maximum distance of 6.6\,kpc before it turned off the main sequence, hence if the IGL were formed from stripped stars we would expect the regions furtherest from any galaxy to be reddest. Fig.~\ref{color} shows that the opposite is true: the regions furthest from a galaxy are bluer than regions close to a galaxy. This suggests that the stars that comprise the IGL have not passively evolved, hence star formation occurs outside the galaxies in a $>60$\,kpc halo surrounding the central radio galaxy, encompassing the Spiderweb complex. An alternative possibility is that stars are stripped from the satellite galaxies together with molecular clouds which then form stars in the halo, however, this cannot be differentiated from star formation fueled by a gas reservoir in the halo. The similarity between the colors of the IGL just outside the galaxies and the galaxies themselves suggest some mixing of stellar populations or a common origin.

There are a number of absorption features in the \lya\ lines shown in Fig.~\ref{spectra} which implies that cool dense gas coexists with the ionized gas in the halo surrounding the radio galaxy. These cool dense regions can be the fuel for the star formation.
The rest-frame UV continuum flux at 1500\AA\ measured by the \gband\ band is a good tracer of star formation and can be converted to a star formation rate through
\begin{equation}
{\rm SFR ( M_{\odot}~yr^{-1})}= \frac{L_{1500\AA}}{8 \times 10^{27}} {\rm (erg\,s^{-1} Hz^{-1})}
\end{equation}
for a Salpeter initial mass function \citep{Madau1998}.  The total \gband\ band flux  converts to a star formation rate of 130$\pm13$\,M$_{\odot}$\,yr$^{-1}$ in an 4243\,kpc$^{2}$ region with 57$\pm8$\,M$_{\odot}$\,yr$^{-1}$ of stars forming in the region between the galaxies (approximately 85 per cent of the area with a surface brightness $>27$\,mag\,arcsec$^{-2}$ in \iband, i.e.~the total area of the galaxies and the IGL). Therefore almost half of the star formation traced by UV light occurs outside any galaxy-like structure.

\subsection{$\beta$--surface brightness relation for the galaxies and intergalactic light} 

\begin{figure*}
\includegraphics[width=2\columnwidth]{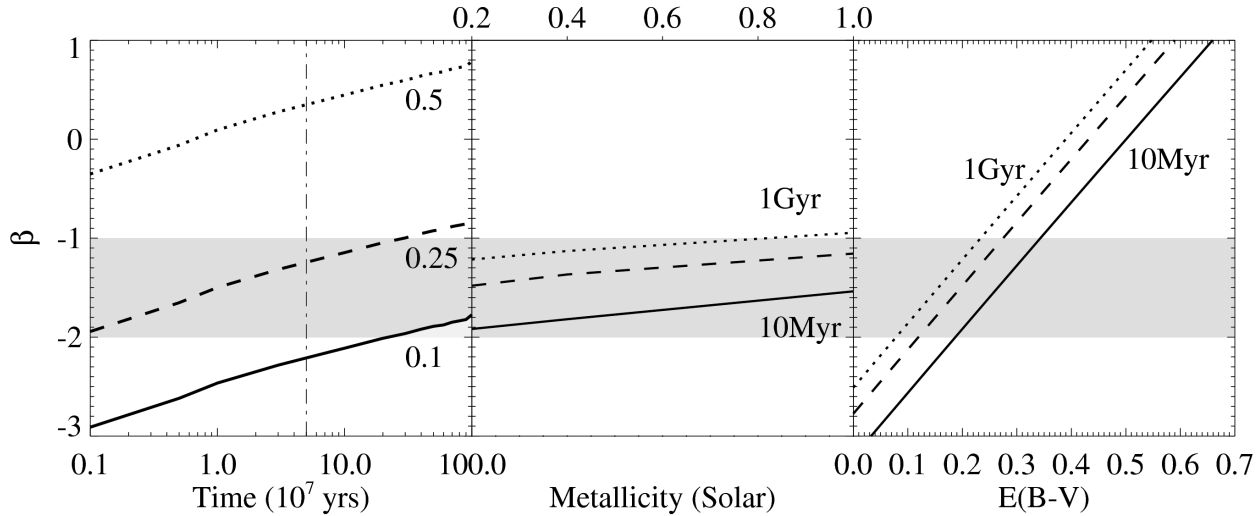}
\vspace{3mm}
\caption{Range of UV slope $\beta$ verses time since start of star burst (in units of $10^7$\,yrs), metallicity (in solar units) and extinction $E(B-V)$. For first panel all lines have a constant metallicity of 0.4\Zsun. The solid line has an extinction  $E(B-V)=0.1$, the dashed line has $E(B-V)=0.25$, and the dotted line has $E(B-V)=0.5$. The vertical dot-dash line marks the minimum time that a stellar cluster which formed on the outskirts of the halo would take to reach the centre. The lines in the middle and right-hand panel have a constant extinction of $E(B-V)=0.2$.  The three different line-styles are cuts at different times since the start of the star formation. The solid line is 10\,Myrs, the dashed line is 200\,Myrs, and the dotted line is 1Gyr. The shaded region indicates the range of $\beta$ observed in the majority of the IGL. \label{beta_all}}\end{figure*}

Having determined that the IGL originates from in-situ star formation we can now examine how $\beta$ depends on metallicity, dust extinction and time since the starburst onset using the stellar population synthesis models {\sc galaxev} of \citet{BC2003}. We calculate $\beta$ as we have defined it in equation \ref{beta}, using the $G$ and $I$ magnitudes outputted from {\sc galaxev}. Fig.~\ref{beta_all} displays the results of the {\sc galaxev} modeling in which we determined the affects of variations in dust, metallicity and age. A star cluster formed at a radius of 30\,kpc from the radio galaxy will be pulled by the gravity of the $10^{12}$\Msun\ radio galaxy, reaching the centre of the halo at $\sim5\times10^7$\,yrs after it has formed (assuming it has no angular momentum). This time is much greater than the lifetime of the OB stars which produce the UV radiation. Since the whole halo is UV bright the entire halo must be continuously forming stars. We therefore use a model of a continuous star formation lasting for 1\,Gyr. In all models we use the Chabrier initial mass function \citep{Chabrier2003}.

The left panel of Fig.~\ref{beta_all} displays the relation between $\beta$ and time since the onset of starburst. All plotted models have a metallicity of 0.4\Zsun and three different extinctions are plotted: $E(B-V)=0.1$ (solid line), $E(B-V)=0.25$ (dashed line), and  $E(B-V)=0.5$ (dotted line). The fainter dash-dot line marks the minimum time that a stellar cluster which formed on the outskirts of the halo would take to reach the centre. If the stellar cluster has any angular momentum this time will be longer. The range in the UV slope that is observed may be produced by a constantly star forming stellar population if the stars at the centre are almost a Gyr old whilst the stars at the edge of the halo are only a Myr old, and the extinction is $E(B-V)>0.2$. Therefore the gradient in $\beta$ can be caused by an ageing of the stellar population if the star formation occurs on the outskirts of the halo, and the stars move inwards as they age. If the star formation rate has decreased over this period the range in $\beta$ would be much larger than plotted in Fig.~\ref{beta_all}, since the older populations would have a greater flux  contribution from the older stellar population. Therefore the range in time may be less than 1\,Gyr. The middle panel of Fig.~\ref{beta_all} shows that the range of metallicity between 0.2-1\Zsun\ results in a change in $\beta$ of $\sim0.4$.  Therefore metallicity variations in a young stellar population cannot produce the radial gradient in $\beta$ for either the galaxies or the intergalactic light. The right-hand panel  of Fig.~\ref{beta_all} shows $\beta$ verses extinction. A range in extinction of $\Delta E(B-V)\sim0.16$ is enough to produce the range in $\beta$ observed for the IGL and just over double this amount will produce the entire range of $\beta$ observed in the galaxies as well as the IGL. 

The IGL of the Spiderweb system is redder than an unobscured starburst, the left-hand panel of Fig.~\ref{beta_all} shows that a minimum $E(B-V)\sim0.1$ is required. Correcting the IGL for internal absorption assuming an $E(B-V)\sim0.1$ increases the IGL star formation rate to over 140\Msunpyr and more than 325\Msunpyr\ for the whole Spiderweb system. The range in observed $\beta$ may be produced by extinction or an aging stellar  population, but the current data is unable to differentiate between these possibilities.

\subsection{Summary of results}
44\%$\pm$12\% of the rest-frame FUV light within 8\arcsec\ (67\,kpc) of the radio galaxy exists as intergalactic light (IGL) that is situated between the radio galaxy and the satellite galaxies. In general the colour of the UV light of the galaxies is very similar to that of the surrounding IGL, however, on the average colour of the IGL is bluer than the average colour of the galaxies, and there is a trend for the IGL to become bluer with radius from the radio galaxy. The source of this IGL is not faint cluster galaxies, but is either an intrinsically diffuse source or many unresolved sources that are spread out over a 3168\,kpc$^2$ region. We have ruled out nebular continuum emission, synchrotron, inverse Compton, synchrotron self-Compton emission and scattering of either galactic stellar light or quasar light as possible sources of the IGL. We conclude that the source of the IGL is most likely to be a distributed young stellar population situated between the radio galaxy and satellite galaxies, which has a UV-derived star formation rate of 57\Msunpyr. 

Both the galaxies and the IGL show a $\beta$--surface brightness relation which cannot be reproduced in stellar population models by a reasonable range in metallicity, but can be obtained by variations in either age or extinction, or a weighted combination of both. Either the halo is populated by a stellar population with a range of ages from a Myr at the outskirts of the halo to 1\,Gyrs in the centre, and has an extinction of $E(B-V)\sim0.25$. Or, irrespective of the age of the halo population, the extinction decreases from the centre to the outskirts by $\Delta E(B-V)\sim0.16$. Correcting for the minimum extinction observed [$E(B-V)\sim0.1$] increases the intergalactic star formation rate to over 140\Msunpyr. Most of the galaxies must either have a much higher extinction than this, or contain significantly older stellar populations. Both of these models can reproduce the radial profile of the UV color.

\section{Discussion}
\label{discussion}
\subsection{Relation between the halo of star formation and the \lya\ halo}
The spectrum of the \lya\ halo that encompasses the Spiderweb system contains strong C{\sc iv} and He{\sc ii} emission lines, which suggests that a harder ionizing source than stellar photoionization is at work here. However, the UV photons from the hot, young stars can contribute to the excitation of the halo and even dominate in some locations. We can estimate the maximum \lya\ flux that can be produced by the total star formation rate of the galaxies and IGL by inverting the relation between H$\alpha$ flux and the star formation rate
\begin{equation}
L_{H\alpha} {\rm (erg\,s^{-1})}={1.27 \times 10^{41}}{\rm SFR ( M_{\odot}~yr^{-1})}
\end{equation}
\citep{Kennicutt1994}.  Assuming case B recombination with no dust absorption (\lya\ /H$\alpha$=11.4) and a total \lya\ luminosity of $2.6\times10^{45}$\ergps \citep{Kurk2003}, this star formation rate can provide only 7\% of the observed \lya\ emission. However the observations are unlikely to be complete and there may exist a fainter component of the IGL that has not been included in this calculation. Therefore 7\% may be considered a lower limit, although it is unlikely that this fraction can increase by more than a factor of 2. This leads to the conclusion that the \lya\ halo is powered by an additional mechanism to star formation. The spatial coincidence of a radio hotspot and the \lya\ peak has sparked speculation that shock heating may be involved. Other mechanisms  for producing the \lya\ emission include AGN photoionization and gravitational cooling radiation. 

\subsection{Low-redshift intracluster light and cD halos}
The low redshift  intracluster light (ICL) is composed of old stars that are thought to have been tidally stripped from merging galaxies or the brightest cluster galaxy (BCG) \citep{Mihos2005} or may have formed from H{\sc i} stripped from the galaxies during encounters.  The colour of the ICL is generally redder than the cluster galaxies and there is evidence for reddening of the ICL with radius from the BCG \citep{Krick2006}. The redder colour of the ICL agrees with the hypothesis that the stars were stripped from the galaxies and then passively evolved, whilst the galaxies continued to form stars. The diffuse UV light from MRC1138-262  is generally bluer than the satellite galaxies and the light becomes bluer with radius. This is inconsistent with the theory that the IGL stars were stripped from the satellite galaxies. It therefore appears that the these these phenomena have very different formation scenarios. The young stars in the halo of MRC\,1138-262 cannot be the high-redshift analogues to the ICL observed at the cores of many low-redshift clusters, e.g.~Virgo \citep{Mihos2005}. However, the high-redshift halo stars may eventual become redder than the satellite galaxies if star formation ceases in the halo prior to the satellite galaxies. Therefore the high-redshift diffuse stars may contribute to the ICL at low redshift. 

Many BCGs show extended halos of light beyond the de Vaucouleurs profile \citep{Matthews1964,Schombert1988}. These galaxies, such as M87, are known as cD galaxies. Deep observations of M87 show that the halo light gradually becomes bluer with radius \citep{Liu2005}, similar to the IGL of MRC\,1138-262. However the observations of cD halos vary widely in their conclusions. Both red and blue halos are found \citep{Schombert1988,Mackie1992}, with red \citep{Gonzalez2000} and blue colour gradients \citep{Zibetti2005}. In order for the halo of intergalactic stars to evolve into the halo of cD galaxies the halo must stop forming stars before a redshift of 1 and the population must age at least 5 Gyrs to reproduce the typical red colors of the halo. \citet{Liu2005} measure the envelope mass of M87 to be $3\times10^{11}$\Msun. Assuming the observed star formation rate of 80\Msunpyr\ is constant, it would take almost 4\,Gyr to produce the necessary mass. Therefore there is just enough time for a cD envelope to form around MRC\,1138-262 from the intergalactic star formation, then ages passively to have the same colour as other cD envelopes at low redshift.

\subsection{Origin of the gas reservoir: Cooling flows, cold accretion model, and outflows}
The high star formation rate of the halo requires that a large gas reservoir is present to provide fuel for the star formation. There are a number of theories as to how this gas reservoir may have arisen. As gas falls into the potential well of a dark matter halo, it can be shock heated to the virial temperature, then radiatively cool, condense and form stars. This is the conventional hot mode of gas accretion. However, simulations of \citet{Fardal2001} and \citet{Kervs2005} show that only half of the gas will be shock heated to the virial temperature of the potential well, whereas the rest will have a much cooler temperature  (T$<10^5$\,K) as it accretes onto the galaxy. This mode of accretion is referred to as the cold accretion model \cite[e.g][]{Haiman2000,Fardal2001}. Alternatively the large star formation rate of the radio and satellite galaxies can produce starburst winds, removing gas from the galaxies and placing it in the halo. In this section we examine each theory in turn.

Starburst mass outflow rates are rather uncertain, but various observational techniques roughly agree that the mass outflow rate is comparable to he star formation rate (see \citealt{Heckman2003} and references therein). Therefore a constant star formation rate of $\sim$70\Msunpyr\ from the radio and satellite galaxies will result in an outflow rate of the same magnitude. The star formation efficiency of nearby molecular clouds is less than 10\% \citep{Jorgensen2007}, therefore the current star formation cannot arise from a gas reservoir produced by a previous outflow unless the star formation rate was a factor of 10 larger at higher redshift.

Cooling flows play an important role in the theory of galaxy formation. A cooling flow occurs when the shock heated infalling gas radiatively cools on a timescale longer than the dynamical time \citep{Fabian1994}. A prerequisite of a cooling flow is the presence of a hot thermalized medium. The high Faraday rotation measure of the polarized radio emission from MRC\,1138-262 suggests that there is a high column density with a magnetic field surrounding the galaxy \citep{Pentericci1997}. There have been  X-ray observations of MRC\,1138-262 with both {\it ROSAT} \citep{Carilli1998} and {\it Chandra} \citep{Carilli2002} satellites. Spatially extended X-ray emission has confirmed an additional source other than the central AGN, however, there is no definite evidence for a thermalised  medium: the extended X-ray emission from MRC\,1138-262 could have originated from either inverse Compton emission or thermal emission from an ICM. 

A strong \lya\ emitting halo could be an observational signature of the cold mode of accretion, and MRC\,1138-262 is surrounded by a large \lya\ emitting halo. The star formation rate of the IGL and the galaxies is insufficient to power the \lya\ halo, therefore gravitational cooling radiation from the cold accretion may play a role. However, simulations have shown that  the \lya\ radiation from gravitational cooling is less (or comparable in massive systems) than that produced by star formation \citep{Fardal2001}, therefore cold accretion may be the origin of the halo, but we require an additional power source other than star formation and gravitational cooling radiation.  Currently we cannot make any strong conclusions as to the origin of the gas reservoir, as all three proposed mechanisms cannot be ruled out.

\subsection{Metal enrichment of the intergalactic medium}
The data suggest that massive stars are forming up to 30\,kpc away from the galaxy center. When these stars die they will release metals into their surroundings, and because they are released so far away from the central galaxy the metals are much more likely to escape the gravitational potential and mix with the intergalactic medium. Numerical models show that high redshift massive galaxies are important contributors to the enrichment of the intracluster medium \citep{DeLucia2004} and inclusion of this radially distant star formation may increase their influence. Further mixing may occur due to starburst winds. While the minimum average star formation rate density in the nebula is $\sim57\Msunpyr / \pi(30kpc)^2$ = 0.02\Msunpyrpsqkpc, the star formation rate density may exceed the superwind threshold of 0.1\Msunpyrpsqkpc\ of \citet{Heckman1990} when accounting for dust or in localized regions throughout the halo.

\subsection{Two modes of star formation in massive galaxies}

These observations show that star formation occurs in two modes in the high redshift universe: one in the usual Lyman break galaxy (LBG) clump-like mode on kpc scales, and the other in a diffuse mode over a large region surrounding massive growing galaxies. Almost 50\% of the star formation traced by UV light occurs in this diffuse mode.
There is a large amount of indirect evidence that similar star forming halos surround a number of other massive high-redshift galaxies: sub-mm measurements of a handful of z$\sim2$ radio galaxies show extended hot dust heated by star formation (Stevens et al.~2003; Greve et al.~2006); a number of z$\sim2$ radio galaxies have extended UV low surface brightness regions \citep{Pentericci1999} and slit spectroscopy of the \lya\ halos surrounding a large sample of high redshift radio galaxies show \lya/He{\sc ii} ratios consistent with massive starbursts \citep{Villar-Martin2007}. \lya\ halos similar in size and luminosity to the halos surrounding radio galaxies have been found toward other high redshift sources as well \cite[e.g][]{Steidel2000,Francis2001,Dey2005}. It is believed that these so-called \lya\ blobs are host to largely obscured, massive starbursts with a possible contribution from superwinds and AGN \cite[e.g][]{Bower2004,Colbert2006}. The new observations of the Spiderweb system provide evidence that such extended star formation can indeed occur in connection with massive forming galaxies and \lya\ halos. Long observations are required to detect the UV low surface brightness extended regions, therefore this phenomenon may occur frequently, but would easily be missed in previous observations.

Clearly this mode of star formation does not continue uninhibited to low redshift as we do not observe much star formation in low redshift brightest cluster galaxies, especially outside the central few kpc. A number of mechanisms may halt star formation in the halo: the supply of gas may be exhausted, or a feedback mechanism from the star formation or the central AGN may blow away the reservoir of gas. At low redshift  heating of the intracluster medium by the AGN is believed to be vitally important in preventing a cooling flow \citep{Binney1995,Churazov2001}. In a related problem, some cosmological simulations invoke AGN heating to halt star formation in massive galaxies \citep{Croton2006}. Although a radio source is present in MRC\,1138-262, the observations of the IGL show that it is unable to prevent the gas from cooling and forming stars. The radio jets may not be able to couple to the surrounding medium as efficiently as it is supposed at low redshift, alternatively if the gas is accreting through a cold mode with a temperature of T$<10^5$\,K the radio phase of the AGN heating may not be powerful enough to prevent the gas from cooling, since the cooling function of gas at temperatures of $10^4-10^5$\,K is much stronger than at hotter temperatures. Additionally, the cold mode of gas accretion declines at low redshift, hence if the gas reservoir is fueled by cold gas accretion, we only expect to observe such halos of star formation at high redshifts. 

\section{Conclusion}
The radio galaxy MRC\,1138-262 is surrounded by a halo of IGL that extends across 60\,kpc. After examining nebular continuum emission, synchrotron, inverse Compton, synchrotron self-Compton emission, scattering, and stripping of stars as possible sources of the IGL, we conclude the most likely origin of the IGL is in-situ star formation.  The minimum star formation rate (i.e. uncorrected for dust) of the IGL is 57$\pm8$\Msunpyr, which is comparable to the total star formation rate derived from the UV luminosity integrated over all the galaxies within 70 kpc of the radio galaxy (and including the radio galaxy itself). Applying a minimum dust correction of $E(B-V)\sim0.1$ imply by the red colours of the the IGL increases the star formation rate of the IGL to 142\Msunpyr, and that of the whole Spiderweb system to more than 325\Msunpyr. We estimate that the total observed star formation rate can produce approximately $\sim$7\% of the \lya\ emission in the halo surrounding the galaxy. The radial colour gradient of the IGL indicates that there is a smooth range in extinction or stellar population age from the outer to the inner parts of the halo, where the inner parts are older or dustier than the outer region. 

While the presence of large quantities of ionized gas found around several remarkable species of high redshift galaxies has been attributed to a number of proposed mechanisms that include AGN feedback, infall, cooling flows, starburst superwinds and mergers, these observations of MRC\,1138-262 show that any successful model should be able to accommodate the mode of extended star formation present in this paper. Our data suggest that the formation of the most massive galaxies is connected with that of their gaseous envelopes and star forming halos, part of which may precede the intracluster light or cD envelopes, or perhaps may contribute to a satellite population. A significant amount of star formation might be occurring in the form of extended low surface brightness features, beyond the typical UV detection limits, as well as in largely obscured extended halos as detected at infrared and sub-mm wavelengths.

\section{Acknowledgments}
NAH and GM acknowledge funding from the Royal Netherlands Academy of Arts and Sciences. JK is supported by the DFG, Sonderforschungsbereich (SFB) 439. We thank the referee for helpful comments, Andrew Zirm and Erica Ellingson for useful discussions, and Michele Cappellari for making his Voronoi 2D binning {\it IDL} programs publicly available at http:$//$www.strw.leidenuniv.nl$/\sim$mcappel$/$idl.
This research has been based on observations made with (1) the NASA/ESA Hubble Space Telescope, obtained at the Space Telescope Science Institute, which is operated by the Association of Universities for Research in Astronomy, Inc., under NASA contract NAS 5-26555. These observations are associated with program 10327. (2) the VLT at ESO, Paranal, Chile, program 64.O-0134(D).

\bibliographystyle{mn2e}\bibliography{mn-jour,1138}

\begin{thebibliography}{}

\bibitem[\protect\citeauthoryear{{Bertin} \& {Arnouts}}{{Bertin} \&
  {Arnouts}}{1996}]{Bertin1996}
{Bertin} E.,  {Arnouts} S.,  1996, \aaps, 117, 393

\bibitem[\protect\citeauthoryear{{Binney} \& {Tabor}}{{Binney} \&
  {Tabor}}{1995}]{Binney1995}
{Binney} J.,  {Tabor} G.,  1995, \mnras, 276, 663

\bibitem[\protect\citeauthoryear{{Blakeslee}, {Anderson}, {Meurer},
  {Ben{\'{\i}}tez} \& {Magee}}{{Blakeslee} et~al.}{2003}]{Blakeslee2003}
{Blakeslee} J.~P.,  {Anderson} K.~R.,  {Meurer} G.~R.,  {Ben{\'{\i}}tez} N.,
  {Magee} D.,  2003, in {Payne} H.~E.,  {Jedrzejewski} R.~I.,   {Hook} R.~N.,
  eds, Astronomical Data Analysis Software and Systems XII Vol.~295 of
  Astronomical Society of the Pacific Conference Series, {An Automatic Image
  Reduction Pipeline for the Advanced Camera for Surveys}.
pp 257--+

\bibitem[\protect\citeauthoryear{{Bower}, {Morris}, {Bacon}, {Wilman},
  {Sullivan}, {Chapman}, {Davies}, {de Zeeuw} \& {Emsellem}}{{Bower}
  et~al.}{2004}]{Bower2004}
{Bower} R.~G.,  {Morris} S.~L.,  {Bacon} R.,  {Wilman} R.~J.,  {Sullivan} M.,
  {Chapman} S.,  {Davies} R.~L.,  {de Zeeuw} P.~T.,    {Emsellem} E.,  2004,
  \mnras, 351, 63

\bibitem[\protect\citeauthoryear{{Brown} \& {Mathews}}{{Brown} \&
  {Mathews}}{1970}]{Brown1970}
{Brown} R.~L.,  {Mathews} W.~G.,  1970, \apj, 160, 939

\bibitem[\protect\citeauthoryear{{Bruzual} \& {Charlot}}{{Bruzual} \&
  {Charlot}}{2003}]{BC2003}
{Bruzual} G.,  {Charlot} S.,  2003, \mnras, 344, 1000

\bibitem[\protect\citeauthoryear{{Cappellari} \& {Copin}}{{Cappellari} \&
  {Copin}}{2003}]{Cappellari2003}
{Cappellari} M.,  {Copin} Y.,  2003, \mnras, 342, 345

\bibitem[\protect\citeauthoryear{{Carilli}, {Harris}, {Pentericci},
  {Rottergering}, {Miley} \& {Bremer}}{{Carilli} et~al.}{1998}]{Carilli1998}
{Carilli} C.~L.,  {Harris} D.~E.,  {Pentericci} L.,  {Rottergering} H.~J.~A.,
  {Miley} G.~K.,    {Bremer} M.~N.,  1998, \apjl, 494, L143+

\bibitem[\protect\citeauthoryear{{Carilli}, {Harris}, {Pentericci},
  {R{\"o}ttgering}, {Miley}, {Kurk} \& {van Breugel}}{{Carilli}
  et~al.}{2002}]{Carilli2002}
{Carilli} C.~L.,  {Harris} D.~E.,  {Pentericci} L.,  {R{\"o}ttgering} H.~J.~A.,
   {Miley} G.~K.,  {Kurk} J.~D.,    {van Breugel} W.,  2002, \apj, 567, 781

\bibitem[\protect\citeauthoryear{{Carilli}, {Roettgering}, {van Ojik}, {Miley}
  \& {van Breugel}}{{Carilli} et~al.}{1997}]{Carilli1997}
{Carilli} C.~L.,  {Roettgering} H.~J.~A.,  {van Ojik} R.,  {Miley} G.~K.,
  {van Breugel} W.~J.~M.,  1997, \apjs, 109, 1

\bibitem[\protect\citeauthoryear{{Chabrier}}{{Chabrier}}{2003}]{Chabrier2003}
{Chabrier} G.,  2003, \pasp, 115, 763

\bibitem[\protect\citeauthoryear{{Churazov}, {Br{\"u}ggen}, {Kaiser},
  {B{\"o}hringer} \& {Forman}}{{Churazov} et~al.}{2001}]{Churazov2001}
{Churazov} E.,  {Br{\"u}ggen} M.,  {Kaiser} C.~R.,  {B{\"o}hringer} H.,
  {Forman} W.,  2001, \apj, 554, 261

\bibitem[\protect\citeauthoryear{{Cimatti}, {di Serego-Alighieri}, {Fosbury},
  {Salvati} \& {Taylor}}{{Cimatti} et~al.}{1993}]{Cimatti1993}
{Cimatti} A.,  {di Serego-Alighieri} S.,  {Fosbury} R.~A.~E.,  {Salvati} M.,
  {Taylor} D.,  1993, \mnras, 264, 421

\bibitem[\protect\citeauthoryear{{Colbert}, {Teplitz}, {Francis}, {Palunas},
  {Williger} \& {Woodgate}}{{Colbert} et~al.}{2006}]{Colbert2006}
{Colbert} J.~W.,  {Teplitz} H.,  {Francis} P.,  {Palunas} P.,  {Williger}
  G.~M.,    {Woodgate} B.,  2006, \apjl, 637, L89

\bibitem[\protect\citeauthoryear{{Croton}, {Springel}, {White}, {De Lucia},
  {Frenk}, {Gao}, {Jenkins}, {Kauffmann}, {Navarro} \& {Yoshida}}{{Croton}
  et~al.}{2006}]{Croton2006}
{Croton} D.~J.,  {Springel} V.,  {White} S.~D.~M.,  {De Lucia} G.,  {Frenk}
  C.~S.,  {Gao} L.,  {Jenkins} A.,  {Kauffmann} G.,  {Navarro} J.~F.,
  {Yoshida} N.,  2006, \mnras, 365, 11

\bibitem[\protect\citeauthoryear{{De Lucia} \& {Blaizot}}{{De Lucia} \&
  {Blaizot}}{2007}]{DeLucia2007}
{De Lucia} G.,  {Blaizot} J.,  2007, \mnras, 375, 2

\bibitem[\protect\citeauthoryear{{De Lucia}, {Kauffmann} \& {White}}{{De Lucia}
  et~al.}{2004}]{DeLucia2004}
{De Lucia} G.,  {Kauffmann} G.,    {White} S.~D.~M.,  2004, \mnras, 349, 1101

\bibitem[\protect\citeauthoryear{{Dey}, {Bian}, {Soifer}, {Brand}, {Brown},
  {Chaffee}, {Le Floc'h}, {Hill}, {Houck}, {Jannuzi}, {Rieke}, {Weedman},
  {Brodwin} \& {Eisenhardt}}{{Dey} et~al.}{2005}]{Dey2005}
{Dey} A.,  {Bian} C.,  {Soifer} B.~T.,  {Brand} K.,  {Brown} M.~J.~I.,
  {Chaffee} F.~H.,  {Le Floc'h} E.,  {Hill} G.,  {Houck} J.~R.,  {Jannuzi}
  B.~T.,  {Rieke} M.,  {Weedman} D.,  {Brodwin} M.,    {Eisenhardt} P.,  2005,
  \apj, 629, 654

\bibitem[\protect\citeauthoryear{{Diehl} \& {Statler}}{{Diehl} \&
  {Statler}}{2006}]{Diehl2006}
{Diehl} S.,  {Statler} T.~S.,  2006, \mnras, 368, 497

\bibitem[\protect\citeauthoryear{{Dubinski}}{{Dubinski}}{1998}]{Dubinski1998}
{Dubinski} J.,  1998, \apj, 502, 141

\bibitem[\protect\citeauthoryear{{Elitzur}, {Nenkova} \&
  {Ivezi{\'c}}}{{Elitzur} et~al.}{2004}]{Elitzur2004}
{Elitzur} M.,  {Nenkova} M.,    {Ivezi{\'c}} Z.,  2004, in {Aalto} S.,
  {Huttemeister} S.,   {Pedlar} A.,  eds, The Neutral ISM in Starburst Galaxies
  Vol.~320 of Astronomical Society of the Pacific Conference Series, {IR
  emission from AGNs}.
pp 242--+

\bibitem[\protect\citeauthoryear{{Fabian}}{{Fabian}}{1989}]{Fabian1989}
{Fabian} A.~C.,  1989, \mnras, 238, 41P

\bibitem[\protect\citeauthoryear{{Fabian}}{{Fabian}}{1994}]{Fabian1994}
{Fabian} A.~C.,  1994, \araa, 32, 277

\bibitem[\protect\citeauthoryear{{Fardal}, {Katz}, {Gardner}, {Hernquist},
  {Weinberg} \& {Dav{\'e}}}{{Fardal} et~al.}{2001}]{Fardal2001}
{Fardal} M.~A.,  {Katz} N.,  {Gardner} J.~P.,  {Hernquist} L.,  {Weinberg}
  D.~H.,    {Dav{\'e}} R.,  2001, \apj, 562, 605

\bibitem[\protect\citeauthoryear{{Francis}, {Williger}, {Collins}, {Palunas},
  {Malumuth}, {Woodgate}, {Teplitz}, {Smette}, {Sutherland}, {Danks}, {Hill},
  {Lindler}, {Kimble}, {Heap} \& {Hutchings}}{{Francis}
  et~al.}{2001}]{Francis2001}
{Francis} P.~J.,  {Williger} G.~M.,  {Collins} N.~R.,  {Palunas} P.,
  {Malumuth} E.~M.,  {Woodgate} B.~E.,  {Teplitz} H.~I.,  {Smette} A.,
  {Sutherland} R.~S.,  {Danks} A.~C.,  {Hill} R.~S.,  {Lindler} D.,  {Kimble}
  R.~A.,  {Heap} S.~R.,    {Hutchings} J.~B.,  2001, \apj, 554, 1001

\bibitem[\protect\citeauthoryear{{Gao}, {Loeb}, {Peebles}, {White} \&
  {Jenkins}}{{Gao} et~al.}{2004}]{Gao2004}
{Gao} L.,  {Loeb} A.,  {Peebles} P.~J.~E.,  {White} S.~D.~M.,    {Jenkins} A.,
  2004, \apj, 614, 17

\bibitem[\protect\citeauthoryear{{Gonzalez}, {Zabludoff}, {Zaritsky} \&
  {Dalcanton}}{{Gonzalez} et~al.}{2000}]{Gonzalez2000}
{Gonzalez} A.~H.,  {Zabludoff} A.~I.,  {Zaritsky} D.,    {Dalcanton} J.~J.,
  2000, \apj, 536, 561

\bibitem[\protect\citeauthoryear{{Greve}, {Ivison} \& {Stevens}}{{Greve}
  et~al.}{2006}]{Greve2006}
{Greve} T.~R.,  {Ivison} R.~J.,    {Stevens} J.~A.,  2006, Astronomische
  Nachrichten, 327, 208

\bibitem[\protect\citeauthoryear{{Haiman}, {Spaans} \& {Quataert}}{{Haiman}
  et~al.}{2000}]{Haiman2000}
{Haiman} Z.,  {Spaans} M.,    {Quataert} E.,  2000, \apjl, 537, L5

\bibitem[\protect\citeauthoryear{{Harris} \& {Grindlay}}{{Harris} \&
  {Grindlay}}{1979}]{Harris1979}
{Harris} D.~E.,  {Grindlay} J.~E.,  1979, \mnras, 188, 25

\bibitem[\protect\citeauthoryear{{Heckman}}{{Heckman}}{2003}]{Heckman2003}
{Heckman} T.~M.,  2003, in {Avila-Reese} V.,  {Firmani} C.,  {Frenk} C.~S.,
  {Allen} C.,  eds, Revista Mexicana de Astronomia y Astrofisica Conference
  Series Vol.~17 of Revista Mexicana de Astronomia y Astrofisica Conference
  Series, {Starburst-Driven Galactic Winds}.
pp 47--55

\bibitem[\protect\citeauthoryear{{Heckman}, {Armus} \& {Miley}}{{Heckman}
  et~al.}{1990}]{Heckman1990}
{Heckman} T.~M.,  {Armus} L.,    {Miley} G.~K.,  1990, \apjs, 74, 833

\bibitem[\protect\citeauthoryear{{J{\o}rgensen}, {Johnstone}, {Kirk} \&
  {Myers}}{{J{\o}rgensen} et~al.}{2007}]{Jorgensen2007}
{J{\o}rgensen} J.~K.,  {Johnstone} D.,  {Kirk} H.,    {Myers} P.~C.,  2007,
  \apj, 656, 293

\bibitem[\protect\citeauthoryear{{Kennicutt} Jr., {Tamblyn} \&
  {Congdon}}{{Kennicutt} et~al.}{1994}]{Kennicutt1994}
{Kennicutt} Jr. R.~C.,  {Tamblyn} P.,    {Congdon} C.~E.,  1994, \apj, 435, 22

\bibitem[\protect\citeauthoryear{{Kere{\v s}}, {Katz}, {Weinberg} \&
  {Dav{\'e}}}{{Kere{\v s}} et~al.}{2005}]{Kervs2005}
{Kere{\v s}} D.,  {Katz} N.,  {Weinberg} D.~H.,    {Dav{\'e}} R.,  2005,
  \mnras, 363, 2

\bibitem[\protect\citeauthoryear{{Krick}, {Bernstein} \& {Pimbblet}}{{Krick}
  et~al.}{2006}]{Krick2006}
{Krick} J.~E.,  {Bernstein} R.~A.,    {Pimbblet} K.~A.,  2006, \aj, 131, 168

\bibitem[\protect\citeauthoryear{{Kurk}}{{Kurk}}{2003}]{Kurk2003}
{Kurk} J.~D.,  2003, PhD thesis, Leiden University, P.O.~Box 9504, 2300 RA
  Leiden, The Netherlands

\bibitem[\protect\citeauthoryear{{Kurk}, {R{\"o}ttgering}, {Pentericci},
  {Miley}, {van Breugel}, {Carilli}, {Ford}, {Heckman}, {McCarthy} \&
  {Moorwood}}{{Kurk} et~al.}{2000}]{Kurk2000}
{Kurk} J.~D.,  {R{\"o}ttgering} H.~J.~A.,  {Pentericci} L.,  {Miley} G.~K.,
  {van Breugel} W.,  {Carilli} C.~L.,  {Ford} H.,  {Heckman} T.,  {McCarthy}
  P.,    {Moorwood} A.,  2000, \aap, 358, L1

\bibitem[\protect\citeauthoryear{{Liu}, {Zhou}, {Ma}, {Wu}, {Yang}, {Li} \&
  {Chen}}{{Liu} et~al.}{2005}]{Liu2005}
{Liu} Y.,  {Zhou} X.,  {Ma} J.,  {Wu} H.,  {Yang} Y.,  {Li} J.,    {Chen} J.,
  2005, \aj, 129, 2628

\bibitem[\protect\citeauthoryear{{Mackie}}{{Mackie}}{1992}]{Mackie1992}
{Mackie} G.,  1992, \apj, 400, 65

\bibitem[\protect\citeauthoryear{{Madau}, {Pozzetti} \& {Dickinson}}{{Madau}
  et~al.}{1998}]{Madau1998}
{Madau} P.,  {Pozzetti} L.,    {Dickinson} M.,  1998, \apj, 498, 106

\bibitem[\protect\citeauthoryear{{Matthews}, {Morgan} \& {Schmidt}}{{Matthews}
  et~al.}{1964}]{Matthews1964}
{Matthews} T.~A.,  {Morgan} W.~W.,    {Schmidt} M.,  1964, \apj, 140, 35

\bibitem[\protect\citeauthoryear{{McCarthy}}{{McCarthy}}{1993}]{McCarthy1993}
{McCarthy} P.~J.,  1993, \araa, 31, 639

\bibitem[\protect\citeauthoryear{{Meurer}, {Heckman} \& {Calzetti}}{{Meurer}
  et~al.}{1999}]{Meurer1999}
{Meurer} G.~R.,  {Heckman} T.~M.,    {Calzetti} D.,  1999, \apj, 521, 64

\bibitem[\protect\citeauthoryear{{Mihos}, {Harding}, {Feldmeier} \&
  {Morrison}}{{Mihos} et~al.}{2005}]{Mihos2005}
{Mihos} J.~C.,  {Harding} P.,  {Feldmeier} J.,    {Morrison} H.,  2005, \apjl,
  631, L41

\bibitem[\protect\citeauthoryear{{Miley}, {Overzier}, {Zirm}, {Ford}, {Kurk},
  {Pentericci}, {Blakeslee}, {Franx}, {Illingworth}, {Postman}, {Rosati},
  {R{\"o}ttgering}, {Venemans} \& {Helder}}{{Miley} et~al.}{2006}]{Miley2006}
{Miley} G.~K.,  {Overzier} R.~A.,  {Zirm} A.~W.,  {Ford} H.~C.,  {Kurk} J.,
  {Pentericci} L.,  {Blakeslee} J.~P.,  {Franx} M.,  {Illingworth} G.~D.,
  {Postman} M.,  {Rosati} P.,  {R{\"o}ttgering} H.~J.~A.,  {Venemans} B.~P.,
  {Helder} E.,  2006, \apjl, 650, L29

\bibitem[\protect\citeauthoryear{{Nesvadba}, {Lehnert}, {Eisenhauer},
  {Gilbert}, {Tecza} \& {Abuter}}{{Nesvadba} et~al.}{2006}]{Nesvadba2006}
{Nesvadba} N.~P.~H.,  {Lehnert} M.~D.,  {Eisenhauer} F.,  {Gilbert} A.,
  {Tecza} M.,    {Abuter} R.,  2006, \apj, 650, 693

\bibitem[\protect\citeauthoryear{{Ouchi}, {Shimasaku}, {Okamura}, {Furusawa},
  {Kashikawa}, {Ota}, {Doi}, {Hamabe}, {Kimura}, {Komiyama}, {Miyazaki},
  {Miyazaki}, {Nakata}, {Sekiguchi}, {Yagi} \& {Yasuda}}{{Ouchi}
  et~al.}{2004}]{Ouchi2004}
{Ouchi} M.,  {Shimasaku} K.,  {Okamura} S.,  {Furusawa} H.,  {Kashikawa} N.,
  {Ota} K.,  {Doi} M.,  {Hamabe} M.,  {Kimura} M.,  {Komiyama} Y.,  {Miyazaki}
  M.,  {Miyazaki} S.,  {Nakata} F.,  {Sekiguchi} M.,  {Yagi} M.,    {Yasuda}
  N.,  2004, \apj, 611, 660

\bibitem[\protect\citeauthoryear{{Pentericci}, {Kurk}, {R{\"o}ttgering},
  {Miley}, {van Breugel}, {Carilli}, {Ford}, {Heckman}, {McCarthy} \&
  {Moorwood}}{{Pentericci} et~al.}{2000}]{Pentericci2000}
{Pentericci} L.,  {Kurk} J.~D.,  {R{\"o}ttgering} H.~J.~A.,  {Miley} G.~K.,
  {van Breugel} W.,  {Carilli} C.~L.,  {Ford} H.,  {Heckman} T.,  {McCarthy}
  P.,    {Moorwood} A.,  2000, \aap, 361, L25

\bibitem[\protect\citeauthoryear{{Pentericci}, {Roettgering}, {Miley},
  {Carilli} \& {McCarthy}}{{Pentericci} et~al.}{1997}]{Pentericci1997}
{Pentericci} L.,  {Roettgering} H.~J.~A.,  {Miley} G.~K.,  {Carilli} C.~L.,
  {McCarthy} P.,  1997, \aap, 326, 580

\bibitem[\protect\citeauthoryear{{Pentericci}, {Roettgering}, {Miley},
  {Spinrad}, {McCarthy}, {van Breugel} \& {Macchetto}}{{Pentericci}
  et~al.}{1998}]{Pentericci1998}
{Pentericci} L.,  {Roettgering} H.~J.~A.,  {Miley} G.~K.,  {Spinrad} H.,
  {McCarthy} P.~J.,  {van Breugel} W.~J.~M.,    {Macchetto} F.,  1998, \apj,
  504, 139

\bibitem[\protect\citeauthoryear{{Pentericci}, {R{\"o}ttgering}, {Miley},
  {McCarthy}, {Spinrad}, {van Breugel} \& {Macchetto}}{{Pentericci}
  et~al.}{1999}]{Pentericci1999}
{Pentericci} L.,  {R{\"o}ttgering} H.~J.~A.,  {Miley} G.~K.,  {McCarthy} P.,
  {Spinrad} H.,  {van Breugel} W.~J.~M.,    {Macchetto} F.,  1999, \aap, 341,
  329

\bibitem[\protect\citeauthoryear{{Reuland}, {van Breugel}, {R{\"o}ttgering},
  {de Vries}, {Stanford}, {Dey}, {Lacy}, {Bland-Hawthorn}, {Dopita} \&
  {Miley}}{{Reuland} et~al.}{2003}]{Reuland2003}
{Reuland} M.,  {van Breugel} W.,  {R{\"o}ttgering} H.,  {de Vries} W.,
  {Stanford} S.~A.,  {Dey} A.,  {Lacy} M.,  {Bland-Hawthorn} J.,  {Dopita} M.,
    {Miley} G.,  2003, \apj, 592, 755

\bibitem[\protect\citeauthoryear{{Rybicki} \& {Lightman}}{{Rybicki} \&
  {Lightman}}{1979}]{Rybicki1979}
{Rybicki} G.~B.,  {Lightman} A.~P.,  1979, {Radiative processes in
  astrophysics}.
New York, Wiley-Interscience, 1979.~393 p.

\bibitem[\protect\citeauthoryear{{Sawicki} \& {Thompson}}{{Sawicki} \&
  {Thompson}}{2006}]{Sawicki2006}
{Sawicki} M.,  {Thompson} D.,  2006, \apj, 642, 653

\bibitem[\protect\citeauthoryear{{Schechter}}{{Schechter}}{1976}]{Schechter197%
6}
{Schechter} P.,  1976, \apj, 203, 297

\bibitem[\protect\citeauthoryear{{Schneider}, {Hall}, {Richards}, {Strauss},
  {Vanden Berk}, {Anderson}, {Brandt}, {Fan}, {Jester}, {Gray}, {Gunn},
  {SubbaRao} \& {et al.}}{{Schneider} et~al.}{2007}]{Schneider2007}
{Schneider} D.~P.,  {Hall} P.~B.,  {Richards} G.~T.,  {Strauss} M.~A.,  {Vanden
  Berk} D.~E.,  {Anderson} S.~F.,  {Brandt} W.~N.,  {Fan} X.,  {Jester} S.,
  {Gray} J.,  {Gunn} J.~E.,  {SubbaRao} M.~U.,    {et al.} 2007, \aj, 134, 102

\bibitem[\protect\citeauthoryear{{Schombert}}{{Schombert}}{1988}]{Schombert198%
8}
{Schombert} J.~M.,  1988, \apj, 328, 475

\bibitem[\protect\citeauthoryear{{Seymour}, {Stern}, {De Breuck}, {Vernet},
  {Rettura}, {Dickinson}, {Dey}, {Eisenhardt}, {Fosbury}, {Lacy}, {McCarthy},
  {Miley}, {Rocca-Volmerange}, {Rottgering}, {Stanford}, {Teplitz} \& {van
  Breugel}}{{Seymour} et~al.}{2007}]{Seymour2007}
{Seymour} N.,  {Stern} D.,  {De Breuck} C.,  {Vernet} J.,  {Rettura} A.,
  {Dickinson} M.,  {Dey} A.,  {Eisenhardt} P.,  {Fosbury} R.,  {Lacy} M.,
  {McCarthy} P.,  {Miley} G.,  {Rocca-Volmerange} B.,  {Rottgering} H.,
  {Stanford} S.~A.,  {Teplitz} H.,    {van Breugel} W.,  2007, ArXiv
  Astrophysics e-prints

\bibitem[\protect\citeauthoryear{{Sirianni}, {Jee}, {Ben{\'{\i}}tez},
  {Blakeslee}, {Martel}, {Meurer}, {Clampin}, {De Marchi}, {Ford}, {Gilliland},
  {Hartig}, {Illingworth}, {Mack} \& {McCann}}{{Sirianni}
  et~al.}{2005}]{Sirianni2005}
{Sirianni} M.,  {Jee} M.~J.,  {Ben{\'{\i}}tez} N.,  {Blakeslee} J.~P.,
  {Martel} A.~R.,  {Meurer} G.,  {Clampin} M.,  {De Marchi} G.,  {Ford} H.~C.,
  {Gilliland} R.,  {Hartig} G.~F.,  {Illingworth} G.~D.,  {Mack} J.,
  {McCann} W.~J.,  2005, \pasp, 117, 1049

\bibitem[\protect\citeauthoryear{{Sol{\'o}rzano-I{\~n}arrea}, {Best},
  {R{\"o}ttgering} \& {Cimatti}}{{Sol{\'o}rzano-I{\~n}arrea}
  et~al.}{2004}]{Solorzano2004}
{Sol{\'o}rzano-I{\~n}arrea} C.,  {Best} P.~N.,  {R{\"o}ttgering} H.~J.~A.,
  {Cimatti} A.,  2004, \mnras, 351, 997

\bibitem[\protect\citeauthoryear{{Spergel}, {Verde}, {Peiris}, {Komatsu},
  {Nolta}, {Bennett}, {Halpern}, {Hinshaw}, {Jarosik}, {Kogut}, {Limon},
  {Meyer}, {Page}, {Tucker}, {Weiland}, {Wollack} \& {Wright}}{{Spergel}
  et~al.}{2003}]{Spergel2003}
{Spergel} D.~N.,  {Verde} L.,  {Peiris} H.~V.,  {Komatsu} E.,  {Nolta} M.~R.,
  {Bennett} C.~L.,  {Halpern} M.,  {Hinshaw} G.,  {Jarosik} N.,  {Kogut} A.,
  {Limon} M.,  {Meyer} S.~S.,  {Page} L.,  {Tucker} G.~S.,  {Weiland} J.~L.,
  {Wollack} E.,    {Wright} E.~L.,  2003, \apjs, 148, 175

\bibitem[\protect\citeauthoryear{{Steidel}, {Adelberger}, {Shapley}, {Pettini},
  {Dickinson} \& {Giavalisco}}{{Steidel} et~al.}{2000}]{Steidel2000}
{Steidel} C.~C.,  {Adelberger} K.~L.,  {Shapley} A.~E.,  {Pettini} M.,
  {Dickinson} M.,    {Giavalisco} M.,  2000, \apj, 532, 170

\bibitem[\protect\citeauthoryear{{Storchi-Bergmann}, {Baldwin} \&
  {Wilson}}{{Storchi-Bergmann} et~al.}{1993}]{Storchi-Bergmann1993}
{Storchi-Bergmann} T.,  {Baldwin} J.~A.,    {Wilson} A.~S.,  1993, \apjl, 410,
  L11

\bibitem[\protect\citeauthoryear{{Tadhunter}, {Dickson}, {Morganti},
  {Robinson}, {Wills}, {Villar-Martin} \& {Hughes}}{{Tadhunter}
  et~al.}{2002}]{Tadhunter2002}
{Tadhunter} C.,  {Dickson} R.,  {Morganti} R.,  {Robinson} T.~G.,  {Wills} K.,
  {Villar-Martin} M.,    {Hughes} M.,  2002, \mnras, 330, 977

\bibitem[\protect\citeauthoryear{{Tadhunter}, {Scarrott}, {Draper} \&
  {Rolph}}{{Tadhunter} et~al.}{1992}]{Tadhunter1992}
{Tadhunter} C.~N.,  {Scarrott} S.~M.,  {Draper} P.,    {Rolph} C.,  1992,
  \mnras, 256, 53P

\bibitem[\protect\citeauthoryear{{Tsvetanov}, {Morse}, {Wilson} \&
  {Cecil}}{{Tsvetanov} et~al.}{1996}]{Tsvetanov1996}
{Tsvetanov} Z.~I.,  {Morse} J.~A.,  {Wilson} A.~S.,    {Cecil} G.,  1996, \apj,
  458, 172

\bibitem[\protect\citeauthoryear{{Venemans}, {R{\"o}ttgering}, {Miley}, {van
  Breugel}, {de Breuck}, {Kurk}, {Pentericci}, {Stanford}, {Overzier}, {Croft}
  \& {Ford}}{{Venemans} et~al.}{2007}]{Venemans2007}
{Venemans} B.~P.,  {R{\"o}ttgering} H.~J.~A.,  {Miley} G.~K.,  {van Breugel}
  W.~J.~M.,  {de Breuck} C.,  {Kurk} J.~D.,  {Pentericci} L.,  {Stanford}
  S.~A.,  {Overzier} R.~A.,  {Croft} S.,    {Ford} H.,  2007, \aap, 461, 823

\bibitem[\protect\citeauthoryear{{Vernet}, {Fosbury}, {Villar-Mart{\'{\i}}n},
  {Cohen}, {Cimatti}, {di Serego Alighieri} \& {Goodrich}}{{Vernet}
  et~al.}{2001}]{Vernet2001}
{Vernet} J.,  {Fosbury} R.~A.~E.,  {Villar-Mart{\'{\i}}n} M.,  {Cohen} M.~H.,
  {Cimatti} A.,  {di Serego Alighieri} S.,    {Goodrich} R.~W.,  2001, \aap,
  366, 7

\bibitem[\protect\citeauthoryear{{Villar-Mart{\'{\i}}n}, {Humphrey}, {De
  Breuck}, {Fosbury}, {Binette} \& {Vernet}}{{Villar-Mart{\'{\i}}n}
  et~al.}{2007}]{Villar-Martin2007}
{Villar-Mart{\'{\i}}n} M.,  {Humphrey} A.,  {De Breuck} C.,  {Fosbury} R.,
  {Binette} L.,    {Vernet} J.,  2007, \mnras, 375, 1299

\bibitem[\protect\citeauthoryear{{Villar-Mart{\'{\i}}n}, {S{\'a}nchez}, {De
  Breuck}, {Peletier}, {Vernet}, {Rettura}, {Seymour}, {Humphrey}, {Stern}, {di
  Serego Alighieri} \& {Fosbury}}{{Villar-Mart{\'{\i}}n}
  et~al.}{2006}]{Villar-Martin2006}
{Villar-Mart{\'{\i}}n} M.,  {S{\'a}nchez} S.~F.,  {De Breuck} C.,  {Peletier}
  R.,  {Vernet} J.,  {Rettura} A.,  {Seymour} N.,  {Humphrey} A.,  {Stern} D.,
  {di Serego Alighieri} S.,    {Fosbury} R.,  2006, \mnras, 366, L1

\bibitem[\protect\citeauthoryear{{Villar-Mart{\'{\i}}n}, {Vernet}, {di Serego
  Alighieri}, {Fosbury}, {Humphrey} \& {Pentericci}}{{Villar-Mart{\'{\i}}n}
  et~al.}{2003}]{Villar-Martin2003}
{Villar-Mart{\'{\i}}n} M.,  {Vernet} J.,  {di Serego Alighieri} S.,  {Fosbury}
  R.,  {Humphrey} A.,    {Pentericci} L.,  2003, \mnras, 346, 273

\bibitem[\protect\citeauthoryear{{Zibetti}, {White}, {Schneider} \&
  {Brinkmann}}{{Zibetti} et~al.}{2005}]{Zibetti2005}
{Zibetti} S.,  {White} S.~D.~M.,  {Schneider} D.~P.,    {Brinkmann} J.,  2005,
  \mnras, 358, 949

\end{thebibliography}

\end{document}